%% file: Revision.tex
\documentclass[aos,preprint]{imsart}

\RequirePackage[OT1]{fontenc}
\RequirePackage{amsthm,amsmath}
\RequirePackage[numbers]{natbib}
\RequirePackage[colorlinks,citecolor=blue,urlcolor=blue]{hyperref}
\startlocaldefs
\numberwithin{equation}{section}
\theoremstyle{plain}
\newtheorem{theorem}{Theorem}[section]

\newtheorem{assumption}{Assumption}[section]
\newtheorem{proposition}{Proposition}[section]
\newtheorem{remark}{Remark}[section]
\newtheorem{corollary}{Corollary}[section]

\usepackage{pdflscape}
\usepackage{geometry}

\endlocaldefs

\usepackage{adjustbox,threeparttable, ccaption, longtable, booktabs, subcaption}
\usepackage{amsfonts}
\usepackage{float}

\begin{document}

\begin{frontmatter}
\title{Estimation of the Number of Components of Non-parametric Multivariate Finite Mixture Models}
\runtitle{Estimation of the number of mixture components}

\begin{aug}
\author{Caleb Kwon\thanksref{m1}\ead[label=e1]{calebkwon@g.harvard.edu}
\ead[label=u1,url]{http://calebkwon.com}}
and
\author{Eric Mbakop\thanksref{m2} \ead[label=e2]{eric.mbakop@ucalgary.edu}}
\runauthor{C. Kwon and E. Mbakop}

\affiliation{Harvard University\thanksmark{m1} and University of Calgary\thanksmark{m2}}

\address{Caleb Kwon\\
Harvard Business School\\
20 N Harvard St.\\
Boston, MA 02613\\
USA\\
\printead{e1}\\
\printead{u1}}

\address{Eric Mbakop\\
Department of Economics\\
University of Calgary\\
2500 University Dr. N.W.\\
Calgary, Alberta T2N 1N4\\
Canada\\
\printead{e2}\\
}
\end{aug}

\begin{abstract}

We propose a novel estimator for the number of mixture components (denoted by $M$) in a non-parametric finite mixture model.  The setting that we consider is one where the analyst has repeated observations of $K\geq2$ variables that are conditionally independent given a finitely supported latent variable with $M$ support points.  Under a mild  assumption on the joint distribution of the observed and latent variables, we show that an integral operator $T$ that is identified from the data has rank equal to $M$. We use this observation, in conjunction with the fact that singular values of operators are stable under perturbations, to propose an estimator of $M$ which essentially consists of a thresholding rule that counts the number of singular values of a consistent estimator of $T$ that are greater than a data-driven threshold. We prove that our estimator of $M$ is consistent, and establish non-asymptotic results which provide finite sample performance guarantees for our estimator. We present a Monte Carlo study which shows that our estimator performs well for samples of moderate size.

\end{abstract}



\end{frontmatter}

\section{Introduction}
Finite mixture models provide a flexible means to model unobserved heterogeneity, and their usage spans across several disciplines including social sciences, medicine, biology and engineering. We refer the reader to Compiani and Kitamura \cite{CK} and McLachlan and Peel \cite{MP} for a discussion of their usage in economics and other disciplines.  

This paper derives a novel estimator for the number of mixture components in a non-parametric finite mixture model. We consider a setting where the analyst observes an i.i.d sample of $K\geq 2$ variables $(X^1,X^2,\cdots,X^K)$ that are assumed to be independent (but not necessarily identically distributed) given some finitely supported latent variable:
   $\Theta$ \newline
 ( $\Theta \in \{1,\cdots,M\}$), i.e,

\begin{equation}
\label{eqnmix}
F(x)=F(x_1,\cdots,x_K)=\sum_{m=1}^{M}P(\Theta=m)\prod_{k=1}^K F_m^k(x_k),
\end{equation}

where $F(x_1,\cdots,x_K)$ denotes the distribution of $X=(X^1,X^2,\cdots,X^K)$ (which is identified from the data), and each \textit{mixture component}  $ \prod_{k=1}^K F_m^k(x_k)$, for $m\in \{1,\cdots,M\}$, represents  the  distribution of $X$ conditional on $\{\Theta=m\}$ (the latter being equal to the product of the marginals under the conditional independence assumption). Here we do not impose any parametric assumption on the distribution of the mixture components. It was shown in Allman, Matias, and Rhodes \cite{AMR} (Theorem 8 and 9) that  if $K\geq 3$ and the  component distributions $\{F_m^k\}_{m=1}^M$ are linearly independent (for each $k\in \{1,\cdots,K\}$),  then the representation \ref{eqnmix} is unique up to swaps of the labels of the mixture components. Hence the joint distribution of $(X,\Theta)$ is identified (up to label swapping) from that of $X$ (see also Hall and Zhou \cite{HZ}, Hettmansperger and Thomas \cite{HT}, and Hall et
al. \cite{HNPE}). Moreover, when $K\geq 2$ and the component distributions are linearly independent,  Kasahara and Shimotsu \cite{HKKS} show that the number of mixture components $M$ is identified. In this paper we provide a new proof of the latter fact. We show that an integral operator $T$ that is identified from the distribution of $X$ has finite rank equal to $M$, and we use this observation to construct a consistent estimator of $M$. Indeed, we prove that a    thresholding rule which essentially counts the number of singular values of a consistent estimator  $\hat{T}$ of $T$ (in the operator norm) greater than a sample size dependent threshold, yields a consistent estimator of $M$. For implementation of our estimator, we provide simple numerical procedures to compute  the singular values of  $\hat{T}$ and the threshold rule. 

 An example of a setting (in economics) where the mixture representation of equation  \ref{eqnmix} arises, is  the study of first and second-price auctions with private values and unobserved heterogeneity. In Hu, McAdams, and Shum \cite{YAS}  (for instance) the authors consider an auction model where bidders' valuations for the auctioned object  are independent given an unobserved heterogeneity $\Theta$. There, $\Theta$ represents characteristics of the auctioned object that are commonly observed by the bidders (and affect their valuations), but which are not observed by the analyst. The conditional independence of  bidders' valuation given $\Theta$ implies that  the bids (which by assumption are observed by the analyst)  are also independent given $\Theta$ and thus satisfy equation \ref{eqnmix}, where $X$ now represents the vector of observed bids. The goal is  to recover the joint distribution of bids and unobserved heterogeneity (all the terms on the right-hand side of equation \ref{eqnmix}) from the distribution of the observed bids (the term of the left-hand side of \ref{eqnmix}). Once the joint distribution of bids and unobserved heterogeneity is identified, standard results from the auction literature (see Hu, McAdams, and Shum \cite{YAS}) can be used to identify the joint distribution of valuations and unobserved heterogeneity, from which the analyst can then perform counter-factual analysis under different auction environments (see also Hu  \cite{HU},  Kasahara and Shimotsu \cite{HKKS1},  Hu, McAdams and Shum \cite{YAS}, An, Hu and Shum \cite{AHM}, Hu and Shum \cite{HS}, Aguirregabiria and Mira \cite{AM}, and Xiao \cite{RX} for other instances in economics where modelling assumptions give rise to the mixture structure of \ref{eqnmix}). Although Hu, McAdams, and Shum \cite{YAS} show that the number of mixture components $M$ is identified in their model, they do not provide a way to estimate it and simply assume it to be known when they estimate the mixture model (Bonhomme, Jochmans, and Robin \cite{BJR1}, Bonhomme, Jochmans, and Robin \cite{BJR2},  Levine, Hunter, and Chauveau \cite{LHC} and Benaglia, Chauveau, and Hunter \cite{BCH}  also provide estimators of the mixture model \ref{eqnmix} under the assumption that $M$ is known).  However, incorrectly specifying the number of mixture components can lead to incorrect inference of the model's parameters. Under the identifying assumption of Hu, McAdams, and Shum \cite{YAS}, our procedure  provides a consistent estimator of the number of mixture components $M$, and can thus be viewed as a first step toward estimating the mixture model \ref{eqnmix}. 

A paper closely related to ours is Kasahara and Shimotsu \cite{HKKS} which studies the identification and estimation of $M$ (or lower bounds on $M$) in Equation \ref{eqnmix}, and  as in this paper, does not impose any parametric restrictions on the distribution of $(X,\Theta)$. There, it is shown that when $K=2$ (for instance), some matrices $P_{\Delta}$ $-$ each one associated to a rectangular partition $\Delta$ of the support of $X=(X^1,X^2)$ $-$ are identified from the distribution of $X$ and have rank at most $M$ (see Section \ref{secKS}). Moreover, under the linear independence assumption, Kasahara and Shimotsu \cite{HKKS} show that  there exist some \textit{good} partitions $\Delta$ for which the associated matrices $P_{\Delta}$ have rank equal to $M$. However, those \textit{good} partitions $\Delta$ for which the matrices $P_{\Delta}$ have rank equal to $M$ depend on the distribution of $X$, and in general (for an arbitrary partition $\Delta$) the rank of $P_{\Delta}$  is only a lower bound on $M$.  The approach of Kasahara and Shimotsu \cite{HKKS}  consists in estimating the rank of $P_{\Delta}$  for a  partition $\Delta$ chosen at the discretion of the analyst. We show below (Section \ref{secKS}) that our approach is closely related to theirs. Indeed, when the components of $X$ are continuous, the matrix $P_{\Delta}$ can be seen as a restriction of our operator $T$ to the finite dimensional subspace of piecewise constant functions on the partition $\Delta$ (see Proposition \ref{prop03} below). 

Our estimator offers many advantages over that of Kasahara and Shimotsu \cite{HKKS}. First, under the linear independence assumption, our estimator always consistently estimates the number of mixture components, whereas that of Kasahara and Shimotsu \cite{HKKS} is in general only consistent to a lower bound on $M$. Hence, to our knowledge, our paper is the first one in the literature to provide a consistent estimator of $M$ under the linear independence assumption. Secondly, when the linear independence assumption does not hold, our estimator is consistent to a lower bound on the number of mixture components which is always at least as large as the lower bound estimated by the method of Kasahara and Shimotsu \cite{HKKS}. Thirdly, we establish non-asymptotic results which provide finite sample performance guarantees for our estimator. In contrast, all the results of Kasahara and Shimotsu \cite{HKKS} are asymptotic in nature, and they do not provide results to assess the finite sample performance of their procedure. Fourthly, unlike the procedure of Kasahara and Shimotsu \cite{HKKS}, our procedure does not require the analyst to have knowledge of a good upper bound $M_0$ on $M$. We show in a simulation study that for moderate sample sizes, the performance of our procedure is comparable to theirs when $M_0$ is \textit{slightly} larger than $M$, and that having $M_0$ much larger or much smaller than $M$ can lead  to a significant reduction in the performance of their procedure. This makes our procedure relatively more appealing in empirical settings where (bounds on) $M$ can plausibly take a wide range of values.

The rest of the paper is organized as follows. In Section \ref{secMI} we introduce the model and provide our main identification results which relate the number of mixture components $M$ to the rank of an integral operator $T$, and in Section \ref{secKS} we discuss the connection between our approach and that of  Kasahara and Shimotsu \cite{HKKS}. Using our identification argument, we provide in Section \ref{secE} an estimator for $M$, and establish some of its statistical properties. Section \ref{secSIM} presents our Monte Carlo study, Section \ref{secA} applies our method to four empirical examples, and Section \ref{secC} concludes. All proofs are provided in Section \ref{appendix} \\
\textbf{Notation}
Given a continuous linear operator $T:{\cal H}_1\rightarrow {\cal H}_2$, where ${\cal H}_1$ and ${\cal H}_2$ are separable Hilbert spaces, we will use $\|\cdot\|_{op}$ to denote the operator norm defined by $\|T\|_{op}:=\sup_{\{u \in {\cal H}_1, \  \|u\|_{{\cal H}_1}=1\}}\|T(u)\|_{{\cal H}_2}$, where $ \|\cdot\|_{{\cal H}_1}$ and $\|\cdot\|_{{\cal H}_2}$ denote the norms associated with the inner product on ${\cal H}_1$ and ${\cal H}_2$ respectively. For $f\in {\cal H}_1$ and $g \in {\cal H}_2$, $g\otimes f$ denotes their tensor product, which is the rank-one operator defined by $g\otimes f:{\cal H}_1\rightarrow {\cal H}_2$ with $g\otimes f (u)=g\langle f,u\rangle_1$, where $u \in {\cal H}_1$ and $\langle\cdot,\cdot\rangle_1$ denotes the inner product on ${\cal H}_1$. When $T$ is compact, we use $\sigma_1(T)\geq \sigma_2(T)\geq \cdots$ to denote the singular values of $T$ in decreasing order (repeated according to their multiplicities). When ${\cal H}_1 = {\cal H}_2={\cal H}$, we use $\|T\|_{HS}$ to denote the Hilbert-Schmidt norm of $T$ defined by $\|T\|_{HS}^2:=\sum_{i=1}^{\infty}\sum_{j=1}^{\infty} \left(\langle e_j,T(e_i)\rangle_{{\cal H}}\right)^2$ where $\{e_i\}_{i=1}^{\infty}$ is an orthonormal basis of ${\cal H}$ (the sum is independent of the choice of the basis).

\section{Model and identification}
\label{secMI}
We consider a K-variate ($K\geq 2$) finite mixture model where the observed random vectors $X^1,\cdots,X^K$ are conditionally independent given some latent variable $\Theta$, as described by equation \ref{eqnmix}. We refer to each $X^k$ ($k=1,\cdots,K$) as a \textit{component of $X$}, and the $X^k$'s can be either discrete or continuous. Our goal is to estimate the number of mixture components $M$ in equation \ref{eqnmix} from an i.i.d sample of $X$. Allman, Matias, and Rhodes \cite{AMR} show that in general, there are distributions $F$ that admit at least two mixture representations as in equation \ref{eqnmix}, with different numbers of mixture components. However, from Proposition 3 of Kasahara and Shimotsu \cite{HKKS} (see also Theorem 8 and 9 of Allman, Matias, and Rhodes \cite{AMR}), when $K\geq 2$ the number of mixture components is identified from the distribution of $X$ ( i.e., all possible representation of the type \ref{eqnmix} have the same number of mixture components) if the conditional distributions of the components of $X$ given $\Theta$ satisfy a \textit{full rank\slash linear independence} condition.  Because this is a key assumption needed to identify $M$, we state it below as a main assumption. We discuss after stating some of our results, how the conclusions change when the full rank condition fails.

\begin{assumption}{\bf (Full rank \slash Linear independence)} 
\label{FR}
There are at least two components $X^{i}$ and $X^{j}$  of $X$ ($i,j\in\{1,\cdots,K\}$) for which the corresponding families of conditional distributions  $\{F^{i}_m\}_{m=1}^M$ and $\{F^{j}_m\}_{m=1}^M$  that appear in equation \ref{eqnmix} are linearly independent.
\end{assumption}
When the components of $X$ are continuous, Assumption \ref{FR} is mild; it is shown in Mbakop \cite{EM} (Proposition 7.4) that it holds generically (see also Proposition 2 in Kasahara and Shimotsu \cite{HKKS}). It requires the distribution of at least two components of $X$ varies sufficiently across the $M$ groups. In fact, in the case of a two components mixtures, two distributions are linearly independent if and only if they are not equal.

%%%%%%%%%%%%%%%%%%%%%%%%%%%%%%%%%%%%%%%%%%%%%%%%%%%%%%%%%%%%%%%%%%%%%%%%%%%%%%%%%%%%%%%%%%%%%%%%%%%%%%%%%%%%%%%%%%%%%%%%%%%%%%%%%%%%

\subsection{The K=2 case}
\label{secK2}
\indent For expositional clarity, we first consider the case where $K=2$, and consider the general case further below. We further assume that the  components of $X$ are continuously distributed, and that $X$ has a density with respect to the Lebesgue measure. The case with discrete components can be handled similarly, and will be discussed further below (see Remark \ref{rem01} and Proposition \ref{propdiscr}). 

Let $d_1$ (resp. $d_2$) denote the dimension of $X_1$ (resp. $X_2$), and set $d=d_1+d_2$, i.e.,  we have $X^1\in \mathbb{R}^{d_1}$, $X^2\in \mathbb{R}^{d_2}$, and $X\in \mathbb{R}^d$.
We assume that the random vector $X$ has a density with respect to the Lebesgue measure on $\mathbb{R}^d$, denoted $f$, which is \textit{square integrable}.
 In what follows, we assume that $d_1=d_2=1$. The higher dimensional cases can be handled similarly. Note that the density $f$ is identified from the data and can be estimated consistently (at some rate) under additional smoothness assumptions. Let $L^2(\mathbb{R})$ denote the Hilbert space of square integrable functions on $\mathbb{R}$, and let the integral operator $T$, $T:L^2(\mathbb{R}) \rightarrow L^2(\mathbb{R})$, be defined by 
\begin{equation}
\label{eqnt}
[T(u)](x_2)=\int_{\mathbb{R}}u(x_1)f(x_1,x_2)dx_1,
\end{equation}
for any $u\in L^2(\mathbb{R})$. Note that the operator $T$ is identified from the data (since it is entirely determined by the density $f$ which is identified from the data), and equation \ref{eqnmix} implies that $T$ has the following representation:
\begin{equation}
\label{eqnmix1}
T=\sum_{m=1}^M \pi_m f^2_m\otimes f^1_m
\end{equation}
where  $f^i_m$ ($i \in \{1,2\}$ and $m \in \{1,\cdots,M\}$) denotes the conditional density of $X^i$ given $\Theta=m$, and $\pi_m=P(\Theta=m)$. \\
\indent The following proposition shows that in general, the operator $T$ has rank (defined as the dimension of the range of $T$) less than or equal to $M$. Moreover, when Assumption \ref{FR} holds, the operator $T$ has rank equal to $M$, and the number of mixture components is identified. The identification of the number of mixture components under Assumption \ref{FR}  was already established in Kasahara and Shimotsu \cite{HKKS} (see Proposition 3 (a)). Besides providing an alternative proof of the identification of $M$, this proposition is useful as it relates $M$ to the rank of the operator $T$$-$a fact which we exploit to estimate $M$ (or a lower bound on $M$). The content of the proposition is similar in spirit to that of Lemma 10 of Allman, Matias, and Rhode \cite{AMR}, and a proof is provided in Section \ref{appendix}

\begin{proposition}
\label{prop01}
Suppose that the distribution of $X=(X^1,X^2)$ satisfies a mixture representation of the form given by equation \ref{eqnmix}. Then we have  $rank(T)\leq M$. Moreover, if Assumption \ref{FR} holds, then $rank(T)=M$.
\end{proposition}
As a consequence of Proposition \ref{prop01}, the operator $T$ is compact, and it admits a singular value decomposition (see Theorem 15.16 in Kress  \cite{KR}) of the form:
\begin{equation}
\label{eqntens}
T=\sum_{m=1}^{rank(T)} \sigma_m\  v_m \otimes u_m.
\end{equation}
Here $\{u_m\}_{m=1}^{rank(T)}$ forms an orthonormal basis for the orthogonal complement (with respect to the inner product on $L^2(\mathbb{R})$) to the null space of $T$, $\{v_m\}_{m=1}^{rank(T)}$ forms an orthonormal basis for the range of $T$, and $\{\sigma_m\}_{m=1}^{rank(T)}$ denote the singular values of $T$ which are strictly positive. We exploit this singular value decomposition further below to construct an estimator for $M$ or a lower bound on $M$, depending on whether or not we maintain Assumption \ref{FR}.\\

We now introduce a family of operators $\{T_h\}_{(h\geq 0)}$, $T_h : L^2(\mathbb{R}) \rightarrow L^2(\mathbb{R}) $, which can be thought of as \textit{regularizations} of the operator $T$, and which are defined by:
\begin{equation}
\label{eqnth}
[T_h(u)](x_2)=\int_{\mathbb{R}}u(x_1)f_h(x_1,x_2)dx_1,
\end{equation}
for any $u\in L^2(\mathbb{R})$, and with the function $f_h$  denoting the convolution of the density $f$ with a ``product kernel":
\begin{equation}
\label{eqnfh}
f_h(x_1,x_2)=\int_{\mathbb{R}^2}f(u,v)K_h(x_1-u)K_h(x_2-v)dudv.
\end{equation}
Here $K_h(\cdot)=(1/h)K(\cdot/h)$, where $K$ is some density function (or kernel function in general) on $\mathbb{R}$ $-$ the density of the standard normal for instance (the dependence of $T_h$ on the choice of the \textit{regularizing} kernel $K$ is left implicit for notational simplicity). We show in Proposition \ref{prop02} below that  $rank(T_h)=rank(T)$ (for all $h>0$) whenever the set where the Fourier transform of $K$ vanishes has Lebesgue measure zero, and thus, the estimation of $rank(T)$ is equivalent to the estimation of $rank(T_h)$ for any $h>0$. As we show in Section \ref{secE}, the main advantage of the operators $T_h$ over the operator $T$, is that they admit consistent \textit{unbiased} estimators, and concentration inequalities can be used to derive bounds on their estimation error. 
\begin{proposition}
\label{prop02}
Let the integral operators $T$ and $T_h$ be defined as in equation \ref{eqnt} and \ref{eqnth}, and let the kernel function $K$, which appears in the definition of the operator $T_h$, be \textit{any} function that belongs to $L^1(\mathbb{R})\cap L^2(\mathbb{R})$ with Fourier Transform that vanishes on a set of Lebesgue measure zero. Then $rank(T_h)=rank(T)$ for any $h>0$, and each operator $T_h$ admits a singular value decomposition
\begin{equation}
\label{eqntensh}
T_h=\sum_{m=1}^{rank(T)} \sigma^h_m  \  v^h_m \otimes u^h_m
\end{equation}
with all the singular values $\{\sigma^h_m\}_{m=1}^{rank(T)}$ strictly positive.
\end{proposition}
\begin{remark} 
Proposition \ref{prop02} remains valid if we only assume that $K\in L^1(\mathbb{R})$, which implies in particular that the operator $T_h$ is well defined as a Hilbert-Schmidt operator (by Young's convolution inequality). The additional restriction $K\in L^2(\mathbb{R})$ is needed for the operators $\hat{T}_h$ (that we introduce in the next section$-$see equation \ref{estt}) to be well defined as Hilbert-Schmidt operators.
\end{remark}

\begin{remark}
\label{remark0}
In general, there is no simple explicit formula which relates the mixture representation of equation \ref{eqnmix1} to the singular value decomposition of equation \ref{eqntens}. However, for some mixture models, both representations coincide and the singular value decomposition is given by the mixture representation. Consider for instance the bi-variate mixture model $X^1=\Theta+U$, $X^2=\Theta+V$, where $\{U,V,\Theta\}$ are independent, $U\sim V \sim uniform([0,1])$ and $Support(\Theta)=\{0,1,2\}$. For this particular example the mixture representation is given by
\begin{equation}
\label{example}
T= \pi_0 \  f^2_0\otimes f^1_0+\pi_1 \  f^2_1\otimes f^1_1+\pi_2  \ f^2_2\otimes f^1_2
\end{equation}
where the densities $f^1_m$ and $f^2_m$ are equal to the density of a $uniform([m,m+1])$, and $\pi_m=P(\Theta=m)$. Since for  $i\in \{1,2\}$  the densities $\{f^i_m\}_{m=0}^2$ have disjoint support, we have $\int f^i_m(x) f^i_{m'}(x)dx=\delta_{mm'}$ ($\delta_{mm'}=0$ if $m\neq m'$ and $\delta_{mm'}=1$ otherwise), and the functions $\{f^i_m\}_{m=0}^2$ are mutually orthogonal with unit ($L^2$) norm. We thus conclude that a singular value decomposition of the operator $T$ is given by 
$$T= \pi_0 \ f^2_0\otimes f^1_0+\pi_1 \  f^2_1\otimes f^1_1+\pi_2 \  f^2_2\otimes f^1_2$$
and the singular values $\{\sigma_m\}$ are given by the proportion of types $\{\pi_m\}$. Note that if for each $i\in \{1,2\}$, the densities $\{f^i_m\}_{m=1}^M$ have disjoint supports but are not necessarily uniformly distributed, then a slight modification of the above argument shows that a singular value decomposition of the operator $T$ is now given by
$$T= \sigma_0  \ \tilde{f}^2_0\otimes \tilde{f}^1_0+\sigma_1 \ \tilde{f}^2_1\otimes \tilde{f}^1_1+\sigma_2  \ \tilde{f}^2_2\otimes \tilde{f}^1_2$$
where $\tilde{f}^i_m=f^i_m/\|f^i_m\|$ and the singular values $\sigma_m$ are given by \linebreak $\sigma_m= \pi_m \|f^1_m\|\|f^2_m\|$, with $\|f\|$ denoting the $L^2$ norm of $f$. In our Monte Carlo study (Section \ref{secSIM}), we will consider designs given by uniform mixtures of the type given by equation \ref{example}. We show in Section \ref{secE} that the performance of our procedure depends on the magnitude of the singular values of $T$, and uniform designs of the type given by equation \ref{example} will have the advantage that their singular values are known exactly. For the other designs that we consider, we will only know that a singular value decomposition exists, but we will not know the exact magnitudes of the singular values. However, we will be able to obtain estimates of the magnitudes of the singular values through simulations. 
\end{remark}

 \indent We now provide some heuristics for our estimation procedure. The full details are given below in Section \ref{secE}. Let  $\{X_i\}_{i=1}^N$ be an i.i.d sample  of $X$,
% (for notational simplicity, we use $X_i$ to denote either the $i^{th}$ observation of the vector $X$ in the sample $\{X_i\}_{i=1}^N$, or the $i^{th}$ component of the vector $X$; although this may raise some confusion, we think that the correct interpretation of $X_i$ will be clear from the context)
 and  let $\hat{T}_h$ be a consistent estimator of $T_h$ in the Hilbert-Schmidt norm (hence in the operator norm) constructed from the sample  $\{X_i\}_{i=1}^N$. Our estimation of $rank(T)$ (equivalently $rank(T_h)$)  hinges on the observation that the singular values of $T_h$ are \textit{stable}. Indeed, by Weyl's inequality for singular values (See Horn and\ Johnson \cite{HJ}, Inequality $3.3.19$ $p.178$), if $\sigma_1(T)\geq\sigma_2(T)\geq \cdots$ denote the singular values of a compact operator $T$ in non-increasing order (repeated according to their multiplicities), then for any compact operators $T$ and $T'$, and for any $i\geq 1$, we have
\begin{equation}
\label{eqnweyl}
|\sigma_i(T)-\sigma_i(T')|\leq \|T-T'\|_{op}.
\end{equation}
 Furthermore, by the Hoffman-Wielandt inequality (see Horn and\ Johnson \cite{HJ}, inequality $3.3.32$ $p.186$ $-$ which is valid in our setting since all the operators that we consider in this paper are of finite rank) we have
\begin{equation}
\label{eqnHW}
\sum_{i\geq 1}|\sigma_i(T)-\sigma_i(T')|^2\leq \|T-T'\|_{HS}^2.
\end{equation}
 As a consequence of inequality \ref{eqnweyl}, if $\hat{\tau}_h(N) =o_p(1)$ is such that $P( \|T_h-\hat{T}_h\|_{op}>\hat{\tau}_h(N))\rightarrow 0$, then a consistent estimator of $rank(T)$ is given by the number of singular values of $\hat{T}_h$ that are larger than $\hat{\tau}_h(N)$, i.e,
\begin{equation}
\label{eqntresh}
\widehat{M}=\#\{i| \sigma_i(\hat{T}_h)\geq \hat{\tau}_h(N)\}.
\end{equation}
Moreover, as a consequence of inequality \ref{eqnHW}, if the threshold $\hat{\tau}_h(N) =o_p(1)$ is now chosen such that $P( \|T_h-\hat{T}_h\|_{HS}>\hat{\tau}_h(N))\rightarrow 0$, then an alternative consistent estimator of $rank(T)$ is given by 
\begin{equation}
\label{eqntresh2}
\widehat{M}=\#\left\{j \  |\left(\sum_{i\geq j}  \sigma_i(\hat{T}_h)^2\right)^{1/2}\geq \hat{\tau}_h(N)\right\}.
\end{equation}
Indeed, setting $R:=rank(T)$ (implying $\sigma_j(T_h)=0$ for all $j>R$),  inequality \ref{eqnHW} implies that for all $j>R$ we have $\left(\sum_{i\geq j}  \sigma_i(\hat{T}_h)^2\right)^{1/2} \leq \|\hat{T}_h-T_h\|_{HS}\leq \hat{\tau}_h(N)$ (with high probability), and that for all $j\leq R$ $\left(\sum_{i\geq j}  \sigma_i(\hat{T}_h)^2\right)^{1/2} \rightarrow \left(\sum_{i\geq j}  \sigma_i(T_h)^2\right)^{1/2}$ which is strictly positive, and thus (with high probability) much larger than the threshold $\hat{\tau}_h(N)$ ($=o_p(1)$). \\
\indent As the Hilbert-Schmidt norm is a Hilbertian norm, we find it easier to control the estimation error of $T_h$ in the Hilbert-Schmidt norm than in the operator norm, and the estimator of $rank(T)$ that we consider in this paper is the one resulting from the Hoffman-Wielandt inequality (equation  \ref{eqntresh2}). We leave the investigation of estimators of the type given by equation \ref{eqntresh} for future research. In Section \ref{secE}, we provide a consistent estimator $\hat{T}_h$ of $T_h$, and a data-driven threshold $\hat{\tau}_h(N)$, for the estimator \ref{eqntresh2}, which converges in probability to zero (as the sample size $N\rightarrow \infty$) and is an upper bound on the estimation error $\|\hat{T}_h-T_h\|_{HS}$ with probability approaching 1 (as $N \rightarrow \infty$). We also provide a simple numerical procedure to compute the singular values of $\hat{T}_h$. 

\begin{remark}
\label{rem00}
As we recall in Section \ref{secKS}, the method of Kasahara and Shimotsu \cite{HKKS} also relates the number of mixture components $M$ to the rank of some operators. Indeed, they show that some matrices $P_{\Delta}$ (defined in equation \ref{eqnks1} below) have rank at most $M$, and their estimation procedure is based on estimating the rank of an empirical analogue of $P_{\Delta}$.
\end{remark}

\begin{remark}
\label{rem01}
A natural extension of the definition of the operator $T$ in \ref{eqnt} to the case with discrete components can be obtained by replacing $f$ in equation \ref{eqnt} by the probability mass function. When both components of $X$ are discrete and of finite support (for instance) the operator $T$ reduces to a matrix, and the estimation of $M$ under Assumption \ref{FR} reduces to the estimation of the rank of a matrix. In the latter setting, the problem becomes essentially finite dimensional, and the method of Kasahara and Shimotsu \cite{HKKS} (like our method) will provide a consistent estimator of $rank(T)$. In fact, as we show below (Proposition \ref{prop03}), the operator $T$ in the discrete case is equal to the matrix $P_{\Delta}$,  with $\Delta$ given by the \textit{finest} partition of the support of $X$. However, when a component of $X$ is continuous, the operator $T$ is a \textit{proper} infinite dimensional operator. In contrast to the approach of Kasahara and Shimotsu \cite{HKKS} that estimates the rank of a restriction of the operator $T$ to a fixed finite dimensional subspace (see Proposition \ref{prop03}) (with the rank of the restriction of $T$ possibly smaller than that of  $T$), the approach of the present paper is fully non-parametric and estimates the rank of $T$ directly.
\end{remark}

In the case where the components of $X$ are discrete but with support that is not necessarily finite, the integral operator $T$ is not well defined, as the density $f$ does not exist, and its natural analogue is a potentially infinite dimensional matrix. However, we show in Proposition \ref{propdiscr} below that the operators $T_h$ (defined in \ref{eqnth} with $f_h$ defined as in \ref{eqn50}) are still well defined integral operators with ranks at most equal to $M$ (and equal to $M$ when Assumption \ref{FR} holds). As a consequence, the estimator of $rank(T_h)$, which we introduce in Section \ref{secE}, remains valid when $X$ has discrete components.

\begin{proposition}
\label{propdiscr}
Suppose that $K=2$ and that the distribution of $X$ satisfies equation \ref{eqnmix}, but does not have a density with respect to the Lebesgue measure. For $h>0$, let $T_h$ and $f_h$ be defined as in equations \ref{eqnth} and \ref{eqn50} respectively, with a kernel $K$ that satisfies the same conditions as in Proposition \ref{prop02}. Then $rank(T_h)$ is the same for all $h>0$, and $rank(T_h)\leq M$. Moreover, if Assumption \ref{FR} holds, then $rank(T_h)=M$.
\end{proposition}

%\begin{remark}
%\label{rem02}
%The requirement that $\{F^{i}_m\}_{m=1}^M$ are linearly independent in Assumption \ref{FR} puts a restriction on the size of the support of the component $X^i$ if it is discrete: it implies that $X^i$ must have at least $M$ support points. 
%\end{remark}

%%%%%%%%%%%%%%%%%%%%%%%%%%%%%%%%%%%%%%%%%%%%%%%%%%%%%%%%%%%%%%%%%%%%%%%%%%%%%%%%%%%%%%%%%%%%%%%%%%%%%%%%%%%%%%%%%%%%%%%%%%%%%%%%%%%%%

\subsection{The general case ($K\geq 2$)} 
\label{secKG2}

We now consider the case where the observed multivariate vector $X$ has more than two components that are conditionally independent, i.e, $X=(X^1,\cdots,X^K)$ with $K\geq 2$, and equation \ref{eqnmix} holds. For each $i \in \{1,\cdots,K\}$, let $d_i$ denote the dimension of the $i^{th}$ component of $X$, i.e.,  $X^i \in \mathbb{R}^{d_i}$,  let ${\cal S}_i \subset \mathbb{R}^{d_i}$ denote the support of $X^i$, and let $L^2({\cal S}_i)$ denote the space of square integrable functions on ${\cal S}_i$. We assume that all the continuous components of $X$ have a joint density with respect to the Lebesgue measure (on the Euclidean space of corresponding dimension). For each $1\leq i<j\leq K$, let $f_{i,j}$ denote the density (or probability mass function in the discrete case) of the pair $(X^i,X^j)$, and let the (associated) integral operator $T_{i,j}:L^2({\cal S}_i) \rightarrow L^2({\cal S}_j)$, which to a square integrable function $u\in L^2({\cal S}_i)$  assigns the element $T_{i,j}(u)\in L^2({\cal S}_j)$ defined by:
\begin{equation}
\label{eqnt2}
[T_{i,j}(u)](x_j)=\int_{{\cal S}_i}u(x_i)f_{i,j}(x_i,x_j)dx_j.
\end{equation}
The following proposition is a straightforward generalization of Proposition \ref{prop01}
\begin{proposition}
\label{prop04}
Suppose that the distribution of $X=(X^1,\cdots,X^K)$ ($K\geq 2$) satisfies a mixture representation of the form \ref{eqnmix}. Then for any $1\leq i<j\leq K$, we have $rank(T_{i,j})\leq M$. Moreover, if Assumption \ref{FR} holds, then $\max_{1\leq i <j \leq K} rank (T_{i,j})=M$, with the maximal rank being achieved by operators $T_{i,j}$ such that each set of distributions $\{F^{i}_m\}_{m=1}^M$ and $\{F^{j}_m\}_{m=1}^M$ is linearly independent.
\end{proposition}
When the components of $X$ are discrete (binary variables for instance), Assumption \ref{FR} becomes restrictive; Assumption \ref{FR} is not satisfied if the components of $X$ have less than $M$ support points (the requirement that the distributions $\{F^{i}_m\}_{m=1}^M$ are linearly independent means that $X^i$ must have at least $M$ support points). However, it is still possible for $M$ to be identified if instead of considering the distributions of pairs of components of $X$, we consider their distributions in groups. To see this, consider for example the case where $\Theta\in \{1,2,3\}$ and $X$ is composed of four identically distributed binary random variables ($X^1,\cdots,X^4$ are supported on $\{0,1\}$) such that for $i \in\{1,2,3,4\}$ we have $F^i_1(0)=1/3$, $F^i_2(0)=1/2$ and $F^i_3(0)=2/3$.  Then Assumption \ref{FR} is not satisfied, as the distributions $\{F^i_m\}_{m=1}^3$ are necessarily linearly dependent. However, if we let $F^{i,j}_m$ denote the conditional distribution of $X^{i,j}:=(X^i,X^j)$ given $\Theta=m$, then it is easy to check that the distributions $\{F^{1,2}_m\}_{m=1}^3$ and $\{F^{3,4}_m\}_{m=1}^3$ are linearly independent, and the operator associated to the random vector $(X^{1,2},X^{3,4})$ has rank equal to $3$. The foregoing suggests the following weaker assumption, which is more adequate when $X$ has discrete components with small supports, and $K>2$ (both assumptions coincide when $K=2$):
\begin{assumption}
\label{FR2}
There exists a partition $\alpha \cup \alpha'$ of $\{1,\cdots,K\}$, with $\alpha$ and $\alpha '$ non-empty, such that the corresponding families of distributions $\{F^{\alpha}_m\}_{m=1}^M$ and $\{F^{\alpha '}_m\}_{m=1}^M$ are linearly independent.
\end{assumption}

Proposition \ref{prop005} below is the natural analogue of Proposition \ref{prop04} when Assumption \ref{FR2} holds (see Corollary 1 in Kasahara and Shimotsu \cite{HKKS} for a similar result), and can be proved by a slight modification of the proof of Proposition \ref{prop01}. Given a partition $\alpha \cup \alpha'$ of $\{1,\cdot,K\}$, with both $\alpha$ and $\alpha '$ non-empty, let $T_{\alpha}$ be the operator associated to the variables $X^{\alpha}$ and $X^{\alpha '}$, defined as in equation \ref{eqnt2}.
\begin{proposition}
\label{prop005}
Suppose that the distribution of $X=(X^1,\cdots,X^K)$ ($K\geq 2$) satisfies a mixture representation of the form \ref{eqnmix}. Then for any partition $\alpha \cup \alpha '$ of $\{1,\cdots,K\}$, $rank(T_{\alpha})\leq M$. Moreover, if Assumption \ref{FR2} holds, then $\max_{\alpha} rank(T_{\alpha})=M$, with the maximum taken with respect to all partitions of $\{1,\cdots,K\}$.
\end{proposition}

%%%%%%%%%%%%%%%%%%%%%%%%%%%%%%%%%%%%%%%%%%%%%%%%%%%%%%%%%%%%%%%%%%%%%%%%%%%%%%%%%%%%%%%%%%%%%%%%%%%%%%%%%%%%%%%%%%%%%%%%%%%%%%%%%%%%

\subsection{Connection to the approach of Kasahara and Shimotsu}
\label{secKS}

In this section, we first give a brief description of the approach proposed by Kasahara and Shimotsu \cite{HKKS}, and then discuss how their procedure relates to ours. As done in Kasahara and Shimotsu \cite{HKKS}, we focus on the case where $K=2$. The case where $K>2$ can be reduced to the case where $K=2$ by considering an aggregation of the components of $X$ (see the paragraph that precedes Assumption \ref{FR2}).\\
\indent Let $X=(X^1,X^2)$ denote a bivariate random vector, with $X^i$ supported on ${\cal S}_i \subset \mathbb{R}^{d_i}$ ($i=1,2$). Let $\Delta=\Delta^1\times \Delta^2$ be a rectangular partition of the support of $X$, with $\Delta^i:=\{\delta_1^i,\cdots,\delta_{|\Delta^i|}^i\}$ forming a partition of ${\cal S}_i$, with each element $\delta_j^i$ being an interval. Given the partition $\Delta$, let $P_{\Delta}\in \mathbb{R}^{|\Delta^1|\times |\Delta^2|}$, denote the matrix with $(i,j)_{th}$ element given by
\begin{equation}
\label{eqnks1}
[P_{\Delta}]_{i,j}=P(X^1 \in \delta_i^1, X^2 \in \delta_j^2).
\end{equation}

 The method of Kasahara and\ Shimotsu \cite{HKKS} hinges on the observation that under the mixture representation of equation \ref{eqnmix}, the matrices $P_{\Delta}$ (for any partition $\Delta$) have rank at most $M$. Indeed, the conditional independence assumption implies that 
$$P(X^1 \in \delta_i^1, X^2 \in \delta_j^2)=\sum_{m=1}^M P(\Theta=m)P(X^1 \in \delta_i^1|\Theta=m)P(X^2 \in \delta_j^2|\Theta=m),$$
and the matrix $P_{\Delta}$ can be written as the sum of $M$ rank 1 matrices as follows
\begin{equation}
\label{matdec}
P_{\Delta}=\sum_{m=1}^M \pi_m  P^1_m \otimes  {P^2_m}
\end{equation}
where $\pi_m=P(\Theta=m)$, $P^1_m$ (with a similar definition for $P^2_m$) is a vector in $\mathbb{R}^{|\Delta^1|}$ with $i^{th}$ element given by $[P^1_m]_i=P(X^1 \in \delta_i^1|\Theta=m)$, and the tensor product $u\otimes v$ here has the simpler interpretation of the vector outer product, i.e, $u\otimes v=uv^T$. As the matrices $P_{\Delta}$ (one for each partition $\Delta$) can be represented as the sum of $M$ rank-one matrices, they each have rank at most $M$. Therefore, any consistent estimator of the rank of $P_{\Delta}$ (for a given partition $\Delta$) will also be a consistent estimator of a lower bound on $M$. The approach of  Kasahara and Shimotsu  \cite{HKKS} essentially consists in constructing such consistent estimators for $rank(P_{\Delta})$. In addition, Kasahara and Shimotsu \cite{HKKS} show that under assumption \ref{FR}, there exists at least one partition $\Delta$ for which $P_{\Delta}$ has rank $M$,  thus  showing that $M$ is identified under Assumption \ref{FR}  (note that such a partition $\Delta$ necessarily satisfies $\min\{|\Delta^1|,|\Delta^2|\}\geq M$). However, the \textit{identifying partitions} $\Delta$ for which $rank(P_{\Delta})=M$ can only be determined from the distribution of $X$, and Kasahara and Shimotsu \cite{HKKS} do not provide a method for choosing/estimating such identifying partitions in finite sample. As a consequence, their approach is in general only consistent to a lower bound on $M$, and is consistent for $M$ only in those cases when the partition $\Delta$ chosen by the analyst happens to satisfy $rank(P_{\Delta})=M$. \\
\indent We now establish the connection between the two approaches. The following proposition shows that  under the assumptions that the supports ${\cal S}_i$ ($i=1,2$) have finite Lebesgue measure, then the matrices $P_{\Delta}$ are simply the restrictions of the integral operator $T$ (equation \ref{eqnt}) to finite dimensional subspaces. A proof is provided in Section \ref{appendix}. When the supports ${\cal S}_i$ have potentially infinite Lebesgue measure, we show in the proof of Proposition \ref{prop03}, using an approximation argument,  that inequality \ref{eqncon2} still holds, although the equation \ref{eqncon1} fails to hold (some of the operators in the identity are not well defined). Before stating the proposition, we first introduce some notation. Given a partition $\Delta=\Delta^1\times \Delta^2$, let ${\cal M}_{\Delta^i}\subset L^2({\cal S}_i)$, for $i\in \{1,2\}$, denote subspaces of piecewise constant functions on the elements of the partition $\Delta^i$, defined by $${\cal M}_{\Delta^i}:=\{\omega \in L^2({\cal S}_i) \ | \ \omega=\sum_{j=1}^{|\Delta^i|} a_j \mathbb{I}_{\delta_{j}^i} , \ \text{with the}\ \ a_i's \in \mathbb{R} \}.$$
Note that the subspace ${\cal M}_{\Delta^i}$, for $i\in \{1,2\}$, has (finite) dimension equal to $|\Delta^i|$, and is non-trivial since ${\cal S}_i$ has finite Lebesgue measure. For $i=1$ or $2$, let $\Gamma_{\Delta^i}:\mathbb{R}^{|\Delta^i|}\rightarrow {\cal M}_{\Delta^i}$, be defined by $\Gamma_{\Delta^i}(a)=\sum_{j=1}^{|\Delta^i|} a_j \mathbb{I}_{\delta_{j}^i} $  (for $a \in \mathbb{R}^{|\Delta^i|}$), and let its adjoint $\Gamma_{\Delta^i}^* : L^2({\cal S}_i)\rightarrow \mathbb{R}^{|\Delta^i|}$ be the operator which to each element $\omega  \in L^2({\cal S}_i)$ assigns the vector $\Gamma_{\Delta^i}^*(\omega) \in \mathbb{R}^{|\Delta^i|}$ with $j^{th}$ component given by   $[\Gamma_{\Delta^i}^*(\omega)]_j=\int_{\delta_j^i}\omega(x_i)dx_i$ (the integral being well defined since ${\cal S}_i$ has finite Lebesgue measure).

\begin{proposition}
\label{prop03}
Suppose that $K=2$ and that the conditional independence (equation \ref{eqnmix}) representation holds. For each partition $\Delta=\Delta^1\times \Delta^2$, we have 
\begin{equation}
\label{eqncon1}
P_{\Delta}^T=\Gamma_{\Delta^2}^* \circ T \circ \Gamma_{\Delta^1}
\end{equation}
where $\circ$ denotes operator composition. As a consequence, for all $\Delta$ we have
\begin{equation}
\label{eqncon2}
rank(P_{\Delta})\leq rank(T).
\end{equation}
Moreover, there exists at least one partition $\Delta$ such that $rank(P_{\Delta})= rank(T)$.
\end{proposition}
\begin{remark}
\label{rem03}
Note that Assumption \ref{FR} is not needed to establish Proposition \ref{prop03}. As a consequence, when Assumption \ref{FR} does not hold, Proposition \ref{prop01} and \ref{prop03} imply that our approach (which estimates the rank of $T$) will be consistent to a lower bound on $M$ that is in general at least as large as the lower bound estimated by the procedure of Kasahara and Shimotsu \cite{HKKS}. Moreover, when linear independence holds, our approach will always be consistent for $M$, whereas that of Kasahara and Shimotsu \cite{HKKS} will in general only be  consistent to a lower bound on $M$. For instance, if the partition $\Delta$ is such that $\max\{|\Delta^1|,|\Delta^2|\}< M$, then any consistent estimator of the rank of $P_{\Delta}$ will be  asymptotically strictly less than $M$ (with probability approaching 1).
\end{remark}

%%%%%%%%%%%%%%%%%%%%%%%%%%%%%%%%%%%%%%%%%%%%%%%%%%%%%%%%%%%%%%%%%%%%%%%%%%%%%%%%%%%%%%%%%%%%%%%%%%%%%%%%%%%%%%%%%%%%%%%%%%%
%%%%%%%%%%%%%%%%%%%%%%%%%%%%%%%%%%%%%%%%%%%%%%%%%%%%%%%%%%%%%%%%%%%%%%%%%%%%%%%%%%%%%%%%%%%%%%%%%%%%%%%%%%%%%%%%%%%%%%%%%%%
%%%%%%%%%%%%%%%%%%%%%%%%%%%%%%%%%%%%%%%%%%%%%%%%%%%%%%%%%%%%%%%%%%%%%%%%%%%%%%%%%%%%%%%%%%%%%%%%%%%%%%%%%%%%%%%%%%%%%%%%%%%

%%%%%%%%%%%%%%%%%%%%%%%%%%%%%%%%%%%%%%%%%%%%%%%%%%%%%%%%%%%%%%%%%%%%%%%%%%%%%%%%%%%%%%%%%%%%%%%%%%%%%%%%%%%%%%%%%%%%%%%%%%%
%%%%%%%%%%%%%%%%%%%%%%%%%%%%%%%%%%%%%%%%%%%%%%%%%%%%%%%%%%%%%%%%%%%%%%%%%%%%%%%%%%%%%%%%%%%%%%%%%%%%%%%%%%%%%%%%%%%%%%%%%%%
%%%%%%%%%%%%%%%%%%%%%%%%%%%%%%%%%%%%%%%%%%%%%%%%%%%%%%%%%%%%%%%%%%%%%%%%%%%%%%%%%%%%%%%%%%%%%%%%%%%%%%%%%%%%%%%%%%%%%%%%%%%

\section{Estimation}
\label{secE}
In the setting of Section \ref{secK2} ($K=2$, $d_1=d_2=1$), we propose in this section an estimator of $rank(T)$ based on an i.i.d sample $\{X_i\}_{i=1}^N$ of $X$. We discuss further below (see Remark \ref{rem004}) how to extend the results to the general setting ($K>2$). The main result of this section is Theorem \ref{theo1} which provides a consistent estimator of $rank(T)$ of the type given by equation \ref{eqntresh2} with a data-driven threshold, and also provides non-asymptotic performance guarantees for our estimates. The main tools used to derive the results of this section are perturbation theory results (Hoffman-Wielandt inequality-\ref{eqnHW}) and concentration inequalities for sums of independent Hilbert space valued random elements (Theorem 3.4 of Pinelis \cite{PIN}$-$see also Lemma 1 and 2 of Smale and Zhou \cite{SZ}). Our approach is similar to that taken in Koltchinskii and Gine \cite{KG}, Zwald and Blanchard \cite{ZB}, Blanchard, Bousquet, and Zwald \cite{BBZ} and Rosasco, Belkin, and De Vito \cite{RBD}, who also combine perturbation theory results and concentration inequalities to study spectral properties of estimates of integral operators.

 The estimator $\widehat{M}$ that we propose is based on a consistent estimator $\hat{T}_h$ of $T_h$. From Proposition \ref{prop02}, the operator $T$ and the operators $T_h$ ($h>0$) have the same rank. However, as we show below (Proposition \ref{prop50} and Proposition \ref{prop1}), one main advantage of using the operators $T_h$'s (instead of the operator $T$) to estimate $rank(T)$ is that  the operators $T_h$ can be estimated \textit{without bias} and concentration inequalities readily yield simple parametric ($\sqrt{N}$-rate) data-driven bounds on their estimation errors $\|\hat{T_h}-T_h\|_{HS}$. By contrast, the estimation of the operator $T$ necessarily involves a bias term, which may converge to zero at a very slow non-parametric rate, unless the density $f$ is sufficiently smooth. Moreover, the presence of a bias term makes it difficult to obtain good bounds on $\|\hat{T}-T\|_{HS}$, as bounds on the approximation error $\|T-E\hat{T}\|_{HS}$ necessarily depend on smoothness properties of the density $f$ which may be unknown to the analyst.

We now provide a consistent estimator $\hat{T}_h$ of $T_h$, and derive further below (Proposition \ref{prop50}) a data-driven bound $\hat{\tau}_h(N)$ on the estimation error  $\|\hat{T}_h-T_h\|_{HS}$. Note that the function $f_h$ defined in equation \ref{eqnfh} can be rewritten as
\begin{equation}
\label{eqn50}
f_h(x_1,x_2)=EK_h(x_1-X^1)K_h(x_2-X^2).
\end{equation}
Given an i.i.d sample $\{X_i\}_{i=1}^N$, a natural estimator for the operator $T_h$ is given by:
\begin{equation}
\label{estt}
[\hat{T}_h(w)](x_2)=\int_{\mathbb{R}}w(x_1)\hat{f}_h(x_1,x_2)dx_1,
\end{equation}
for any $w\in L^2(\mathbb{R})$,  with the function $\hat{f}_h$ given by the sample analogue of  equation \ref{eqn50}, i.e,
\begin{equation}
\label{eqn51}
\hat{f}_h(x_1,x_2)=\frac{1}{N} \sum_{i=1}^N K_h(x_1-X^1_{i})K_h(x_2-X^2_{i}).
\end{equation}
Since $E\hat{f}_h(x_1,x_2)=f_h(x_1,x_2)$, we have $E \hat{T}_h=T_h$ (see Blanchard, Bousquet and Zwald \cite{BBZ} for the definition of the expectation of a Hilbert space valued random variable, and note that the random variables $\hat{T}_h$ take their values in the space  of Hilbert-Schmidt operators). For each $x=(x_1,x_2) \in \mathbb{R}^2$ and $h>0$, let $T_{h,x}:L^2 (\mathbb{R}) \rightarrow L^2(\mathbb{R})$ denote the rank-one operator defined by 
\begin{equation}
\label{eqn050}
T_{h,x}= K_h(x_2-\cdot)\otimes K_h(x_1-\cdot).
\end{equation}
The following proposition provides a non-asymptotic data-driven bound on the estimation error $\|\hat{T}_h-T_h\|_{HS}=\|\hat{T}_h-E\hat{T}_h\|_{HS}$, with $\hat{T}_h$ defined as in equation \ref{estt}. As noted above, the main tools that we use to derive bounds on the estimation error are concentration inequalities. The proof of the proposition is provided in Section \ref{appendix}.
%%%%%%%%%%%%%%%%%%%%%%%%%%%%%%%%%%%%%%%%%%%%%%%%%%%%%%%%%%%%%%%%
\begin{proposition}
\label{prop50}

For all $\tau>0$ and $N\geq 2$, we have
\begin{equation}
\label{eqnconc1}
\begin{split}
&P(\|\hat{T}_h-T_h\|_{HS}>\tau)\leq 2 \exp\left\{-\frac{\tau N}{L_h}  \ln\left(1+\frac{\tau L_h}{\sigma_h^2}\right)\right\} \times \\
&  \left(1+\frac{\tau}{L_h}-\frac{\sigma_h^2}{L_h^2}\ln \left(1+\frac{\tau L_h}{\sigma_h^2}\right) \right)^N
\end{split}
\end{equation}
where $L_h=:\sup_{x,x'\in \mathbb{R}^2} \|T_{h,x}-T_{h,x'}\|_{HS}$ and $\sigma_h^2:=E\|T_{h,X}-T_h\|_{HS}^2$. As a consequence, for any $0<\delta<1$, the inequality 
\begin{equation}
\label{eqntau}
\|\hat{T}_h-T_h\|_{HS}\leq \tau_h(N,\delta)
\end{equation}
 holds with probability at least $1-\delta$, where $\tau_h(N,\delta)$ is a value of $\tau$ for which the right-hand side of inequality \ref{eqnconc1} is equal to $\delta$. In particular, for $0<\delta<1/2$, the following inequality holds with probability greater than $1-2\delta$:
\begin{equation}
\label{eqn52}
\begin{aligned}
& \|\hat{T}_h-T_h\|_{HS} \leq  \frac{2 L_h ln(2/\delta))}{N} +\\
 &\sqrt{ \frac{ln(2/\delta)}{N} \left( \frac{1}{N(N-1)}\sum_{i\neq j} \|T_{h,X_i}-T_{h,X_j}\|_{HS}^2+ L_h^2 \sqrt{\frac{ln(1/\delta)}{N}}\right)}
 \end{aligned}
\end{equation}
where $X'$ is an independent copy of $X$.
\end{proposition}
%%%%%%%%%%%%%%%%%%%%%%%%%%%%%%%%%%%%%%%%%%%%%%%%%%%%%%%%%%%%%%%%
\begin{remark} From the proof of Proposition \ref{prop50}, the supremum in the definition of  $L_h$ can be replaced by the supremum over the support of $X$ (instead of all of $\mathbb{R}^2$). We have opted for the supremum over all of $\mathbb{R}^2$ to make $L_h$ distribution free (not dependent on the distribution of $X$). In addition, note that the bound on the right-hand side of inequality \ref{eqn52}  can be computed from the data. Indeed, the quantities $L_h$ and $\|T_{h,X_i}-T_{h,X_j}\|_{HS}^2$ can be computed explicitly (or bounded) as they only depend on the  kernel $K$ and the bandwidth $h$, which are both chosen by the analyst. However, the right-hand side of inequality \ref{eqnconc1} depends on the expectation $\sigma_h^2=E\|T_{h,X}-T_{h}\|_{HS}^2$, and cannot be computed from the data. Although Theorem \ref{theo1} below is established with the threshold given by the right-hand side of inequality \ref{eqn52} (equation \ref{eqn53} below), when we implement our method in Section \ref{secSIM}, we use the threshold suggested by inequality \ref{eqntau}. That is, we solve numerically for the value of $\tau$ for which the right-hand side of inequality \ref{eqnconc1} is equal to $\delta$, where we replace $\sigma_h^2$ by its sample analogue $ \frac{1}{2N(N-1)}\sum_{i\neq j} \|T_{h,X_i}-T_{h,X_j}\|_{HS}^2$ (see Section \ref{secH}). 
\end{remark}
Let $\hat{\tau}_h(N,\delta)$ be defined by
\begin{equation}
\label{eqn53}
\begin{aligned}
\hat{\tau}_h(N,\delta)&:=\frac{2 L_h ln(2/\delta))}{N}+\\
&\sqrt{ \frac{ln(2/\delta)}{N} \left( \frac{1}{N(N-1)}\sum_{i\neq j} \|T_{h,X_i}-T_{h,X_j}\|_{HS}^2+ L_h^2 \sqrt{\frac{ln(1/\delta)}{N}}\right)}
\end{aligned}
\end{equation}
and for each $j \in \{1,\cdots,N\}$, define 
\begin{equation}
\label{eqnr}
r_j(\hat{T}_h):=\sqrt{\sum_{i\geq j} \sigma_i(\hat{T}_h)^2}.
\end{equation}
 Note that $\hat{\tau}_h(N,\delta)=o_P(1)$. \\
\indent The following theorem is the main result of this section, and is a direct consequence of  Proposition \ref{prop50} and Hoffman-Wielandt inequality (equation \ref{eqnHW}).
%%%%%%%%%%%%%%%%%%%%%%%%%%%%%%%%%%%%%%%%%%%%%%%%%%%%%%%

\begin{figure}
\subcaptionbox{Mixture of 3 uniform distributions ($N=500$)}{\includegraphics[width=0.49\textwidth]{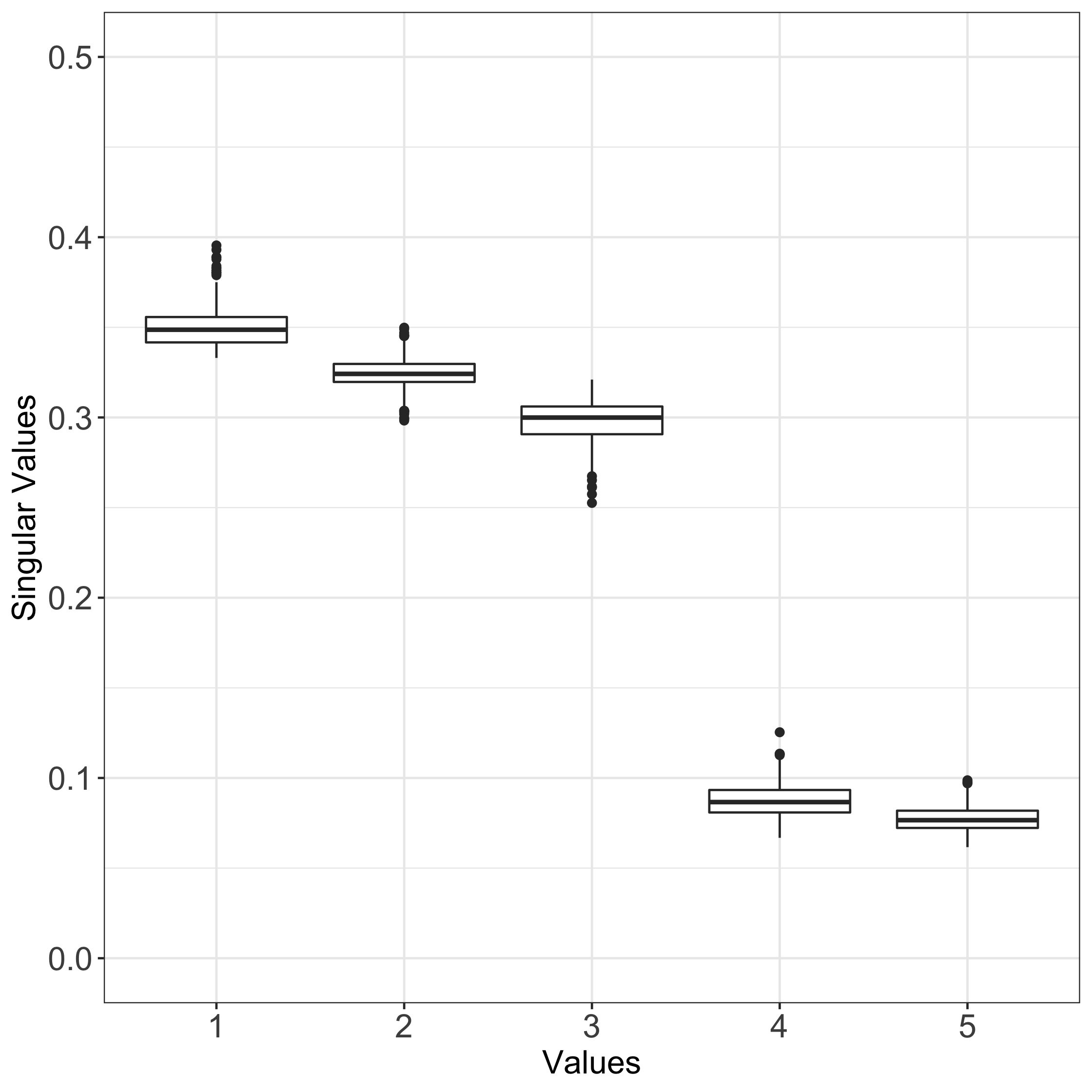}}
\hfill % <-- Seperation
\subcaptionbox{Mixture of 3 uniform distributions ($N=4000$)}{\includegraphics[width=0.49\textwidth]{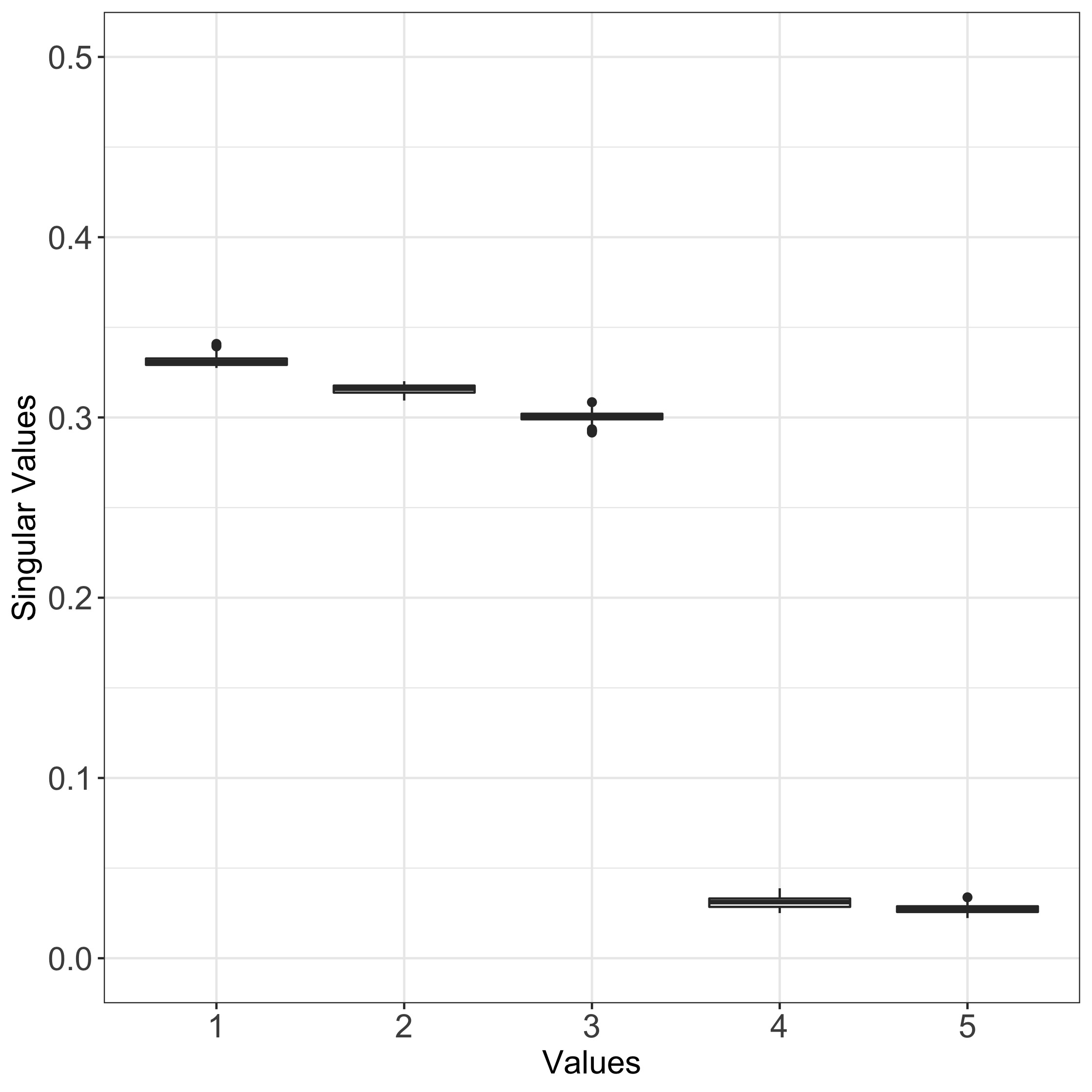}}
\\ % <-- Line break
\subcaptionbox{Mixture of 3 normal distributions ($N=500$)}{\includegraphics[width=0.49\textwidth]{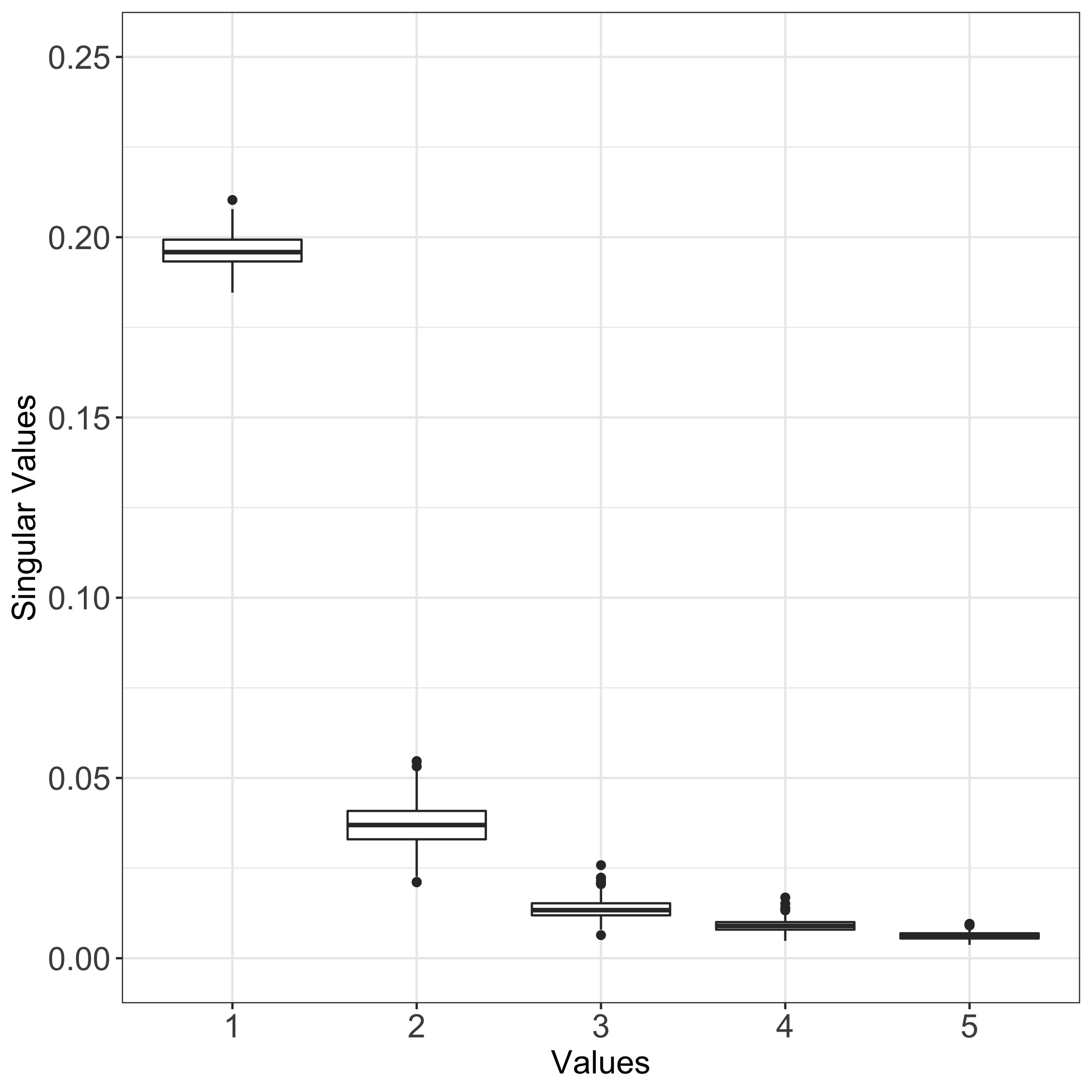}}
\hfill % <-- Seperation
\subcaptionbox{Mixture of 3 normal distributions ($N=4000$)}{\includegraphics[width=0.49\textwidth]{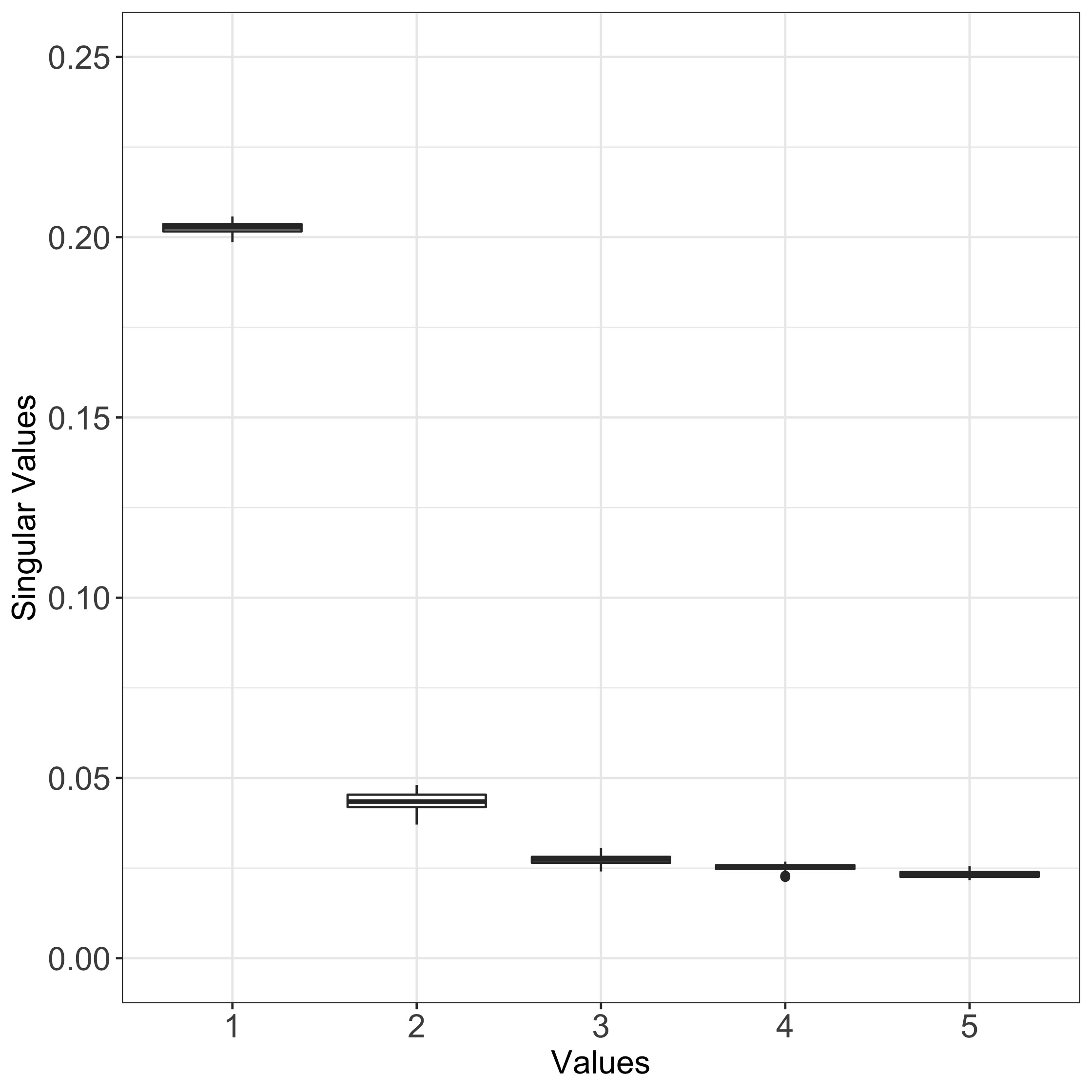}} 
\caption{Box and Whisker plots of the largest five singular values of $\hat{T}$ computed from Equation \ref{eqn01}, with $h=0.05$. Box (a) and (b) corresponds to data generated from a mixture of  3 uniform distributions (design 2 in Section \ref{secSIM}), and Box (c) and (d) to a mixture of 3 normal distributions (design 1 in Section \ref{secSIM}). Note that the three largest singular values of the uniform design are all close to $1/3$ (see Remark \ref{remark0}).}
\label{fig:box_whiskers}
\end{figure}

\begin{figure}[H]
\subcaptionbox{Variable bandwidths $(h)$ }{\includegraphics[width=0.49\textwidth]{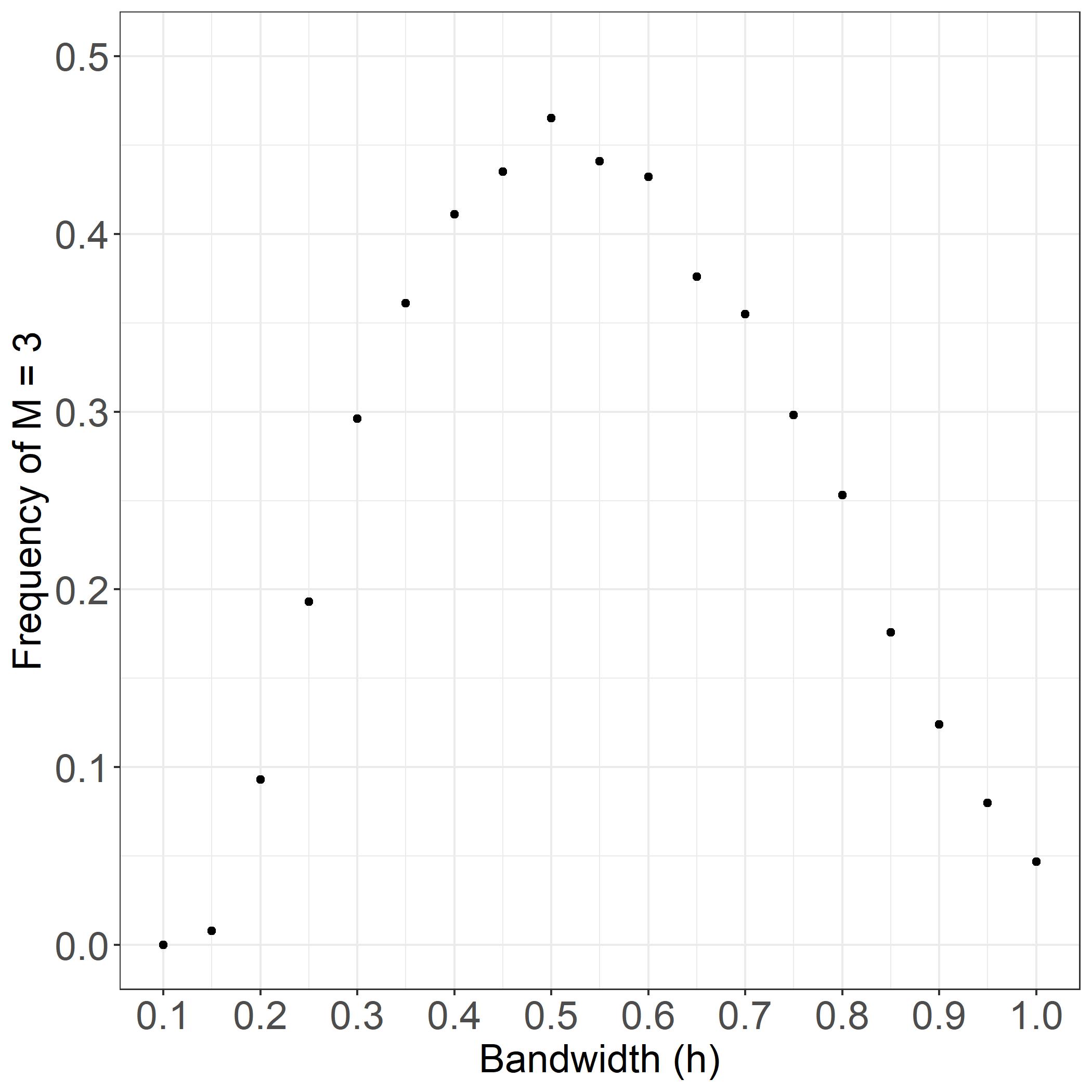}}
\hfill % <-- Seperation
\subcaptionbox{Variable $\delta$ }{\includegraphics[width=0.49\textwidth]{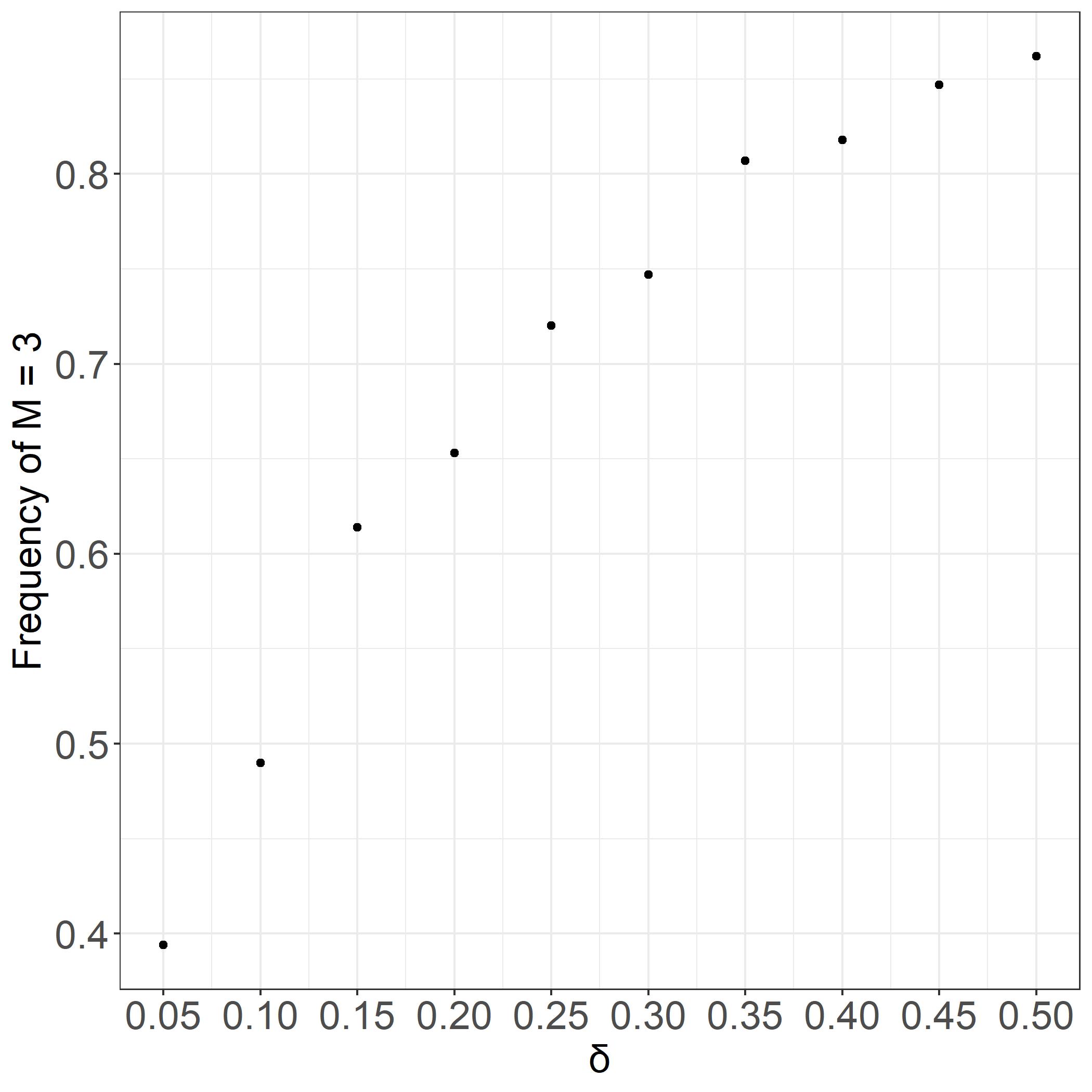}}
\\ % <-- Line break
 
\caption{Selection frequencies of $M =3$ for Design 1 in Section \ref{secSIM} (mixture of 3 Normals), from 1000 Monte Carlo simulations, and a sample size of $N=2000$. In $(a)$, the selection probabilities are given for diffferent values of $h$ when $\delta=0.05$. In $(b)$, the selection probabilities are given for different values of $\delta$, and with the bandwidth chosen according to Silverman's rule.  The value of the bandwidth given by Silverman's rule ranges from 0.421 to 0.504, across the 1000 Monte Carlo simulations.}
\label{fig:bw_delta_DT}
\end{figure}

\begin{theorem}
\label{theo1}
Suppose that the distribution of $X$ satisfies the mixture representation of equation \ref{eqnmix}, and for $h>0$, let $\hat{T}_h$ be defined by equation \ref{estt}. Consider the estimator of $rank(T)$ given by
\begin{equation}
\label{eqnest}
\widehat{M}=\#\{j\ | \  r_j(\hat{T}_h)\geq \hat{\tau}_h(N,\delta)\},
\end{equation}
where $\hat{\tau}_h(N,\delta)$ and $r_j(\hat{T}_h)$ are defined as in equation \ref{eqn53} and \ref{eqnr}. Then, for any $0<\delta<1/2$, we have:
\begin{equation}
\label{eqnest1}
 P(\widehat{M} \leq rank(T)) \geq 1-2\delta,
\end{equation}
\begin{equation}
\label{eqnest2}
P(\{ \sigma_{rank(T)}(T_h) > 2\hat{\tau}_h(N,\delta)\} \cap \{\|\hat{T}_h-T_h\|_{HS}\leq \hat{\tau}_h(N,\delta)\})\leq P(\widehat{M} = rank(T)),
\end{equation}
and
\begin{equation}
\label{eqnest3}
P( \{ \sigma_{rank(T)}(T_h)+\|\hat{T}_h-T_h\|_{HS}<\hat{\tau}_h(N,\delta)\})\leq P(\widehat{M} < rank(T)),
\end{equation}
where $\sigma_{rank(T)}(T_h)$ denotes the smallest nonzero singular value of $T_h$. As a consequence, if $\delta=\delta(N) \rightarrow 0$ and $ln(1/\delta(N))=o(N)$, then $P(\widehat{M} = rank(T)) \rightarrow 1$. Moreover, if Assumption \ref{FR} is satisfied, then $rank(T)=M$ and $\widehat{M}$ is a consistent estimator of $M$.
\end{theorem}
%%%%%%%%%%%%%%%%%%%%%%%%%%%%%%%%%%%%%%%%%%%%%%%%%%%%%%%%%%%%%%
\begin{remark} 
\label{rem001} Inequality \ref{eqnest1} shows that our choice of threshold $ \hat{\tau}_h(N,\delta)$ guarantees that $\widehat{M}$ is a lower bound for $rank(T)$ (and hence for $M$) with probability at least $1-2\delta$ for any $N$. Moreover, Inequality \ref{eqnest2} shows that $\widehat{M}$ is a non-trivial lower bound on $rank(T)$ (the trivial lower bound $\widehat{M} \equiv 1$ also satisfies inequality \ref{eqnest1}), as it implies that for any choice of $\delta$ ($=\delta(N)$) such that $\hat{\tau}_h(N,\delta)=o_P(1)$, we have
\begin{equation}
\liminf_{N\rightarrow \infty} P(\widehat{M} = rank(T))\geq \liminf_{N\rightarrow \infty} (1-2\delta(N)).
\end{equation}
 In particular, inequality \ref{eqnest2} implies that for a given sample size $N$, the estimator $\widehat{M}$ performs well (is equal to $rank(T)$ with high probability) when the smallest nonzero singular value of $T_h$ is well separated from zero relative to the sample size. This is confirmed by our simulation studies; see Figure \ref{fig:box_whiskers} (a) and (b), which correspond to design 2 in Section \ref{secSIM}, where the largest nonzero singular value (third in this case) is well away from zero,  and note the good performance of our method on this design in the simulation study. By contrast, inequality \ref{eqnest3} shows that $\widehat{M}$ underestimates $rank(T)$ with high probability, whenever the smallest non-zero singular value of the operator $T_h$ is close to zero and much smaller than the bound $ \hat{\tau}_h(N,\delta)$ on the estimation error; see Figure \ref{fig:box_whiskers} (c) and (d), which correspond to design 1 in Section \ref{secSIM}, where, as shown by the figures, the smallest nonzero singular value (third in this case) is close to zero, and note the relatively (compared to design 2) worse performance of our method for this design in the simulation study. 
\end{remark}
\begin{remark}
The parameter $\delta$ is chosen by the analyst, and controls the overestimation probability (by inequality \ref{eqnest1} $P(\widehat{M}>rank(T))<2\delta$). Hence, if overestimating $M$ is more of a concern than underestimating $M$, then the analyst should select small values of $\delta$ (e.g. $\delta=0.05$), and when the converse is more desirable, larger values of $\delta$ should be considered (e.g. $\delta=0.4$). Note, however, that as the sample size increases, the parameter $\delta$ must satisfy the conditions $\delta(N)\rightarrow 0$ and $ln(1/\delta(N))=o(N)$, for $\widehat{M}$ to be consistent. Our simulation studies in Section \ref{secSIM} show that for a given value of $\delta$, the probability that our procedure overestimates $M$ is much smaller than the upper bound given by inequality \ref{eqnest1}. For instance, when $\delta=0.05$, our procedure never overestimates $M$, although the latter should be expected to occur with probability close to $2\delta(=0.1)$ if the bound in inequality \ref{eqnest1}  is ``approximately sharp". The slackness in the bound given by inequality \ref{eqnest1} is further illustrated by Table \ref{tabledelta}, where we evaluate the performance of $\widehat{M}$ on Design 1 of Section \ref{secSIM} (mixture of 3 Normals), for various values of $\delta$. Further, even when $\delta=0.5$, $\widehat{M}$ overestimates the true number of components ($M=3$)  less than 1\%, which is much smaller than the upper bound of $2\delta(=1)$ given by inequality \ref{eqnest1}. For this same design, we plot in Figure \ref{fig:bw_delta_DT} (b) the frequencies at which the correct number of components is selected as $\delta$ varies, and note the drastic improvement in performance: $P(\widehat{M}=3)$ increases from 39\% when $\delta=0.05$ to 86\% when $\delta=0.5$. Overall, our simulations seem to indicate that the bound in inequality \ref{eqnest1} is not sharp, and the analyst should select larger values of $\delta$ than suggested by inequality \ref{eqnest1}. In all our simulations in Section \ref{secSIM} and in the empical examples of Section \ref{secA}, we implement our procedure using two values of $\delta$: $\delta \in \{0.05,0.4\}$.

\end{remark}
\begin{remark}
\label{rem003}
The results of  Theorem \ref{theo1} are valid for any choice of bandwidth $h>0$, and as noted in Remark \ref{rem001}, inequality \ref{eqnest2} implies that our procedure correctly estimates $rank(T)$ with high probability, whenever the smallest non-zero singular value of $T_h$ ($\sigma_{rank(T)}(T_h))$ is much larger than the the threshold $ \hat{\tau}_h(N,\delta)$ with high probability. In Proposition \ref{prop1} below, we provide some results that describe the behavior of the singular values of $T_h$ as $h$ tends to zero or infinity. In particular, we show that the smallest nonzero singular values of $T_h$ converge to the smallest nonzero singular value of $T$ as $h\rightarrow 0$, and that the smallest nonzero singular value of $T_h$ tends to zero as $h\rightarrow \infty$.
%; it is not  clear, however, how the singular values of $T_h$ vary with $h$ for intermediate values of h. 
By contrast, for fixed $N$, the threshold $ \hat{\tau}_h(N,\delta)$ tends to zero as $h\rightarrow \infty$, and tends to infinity as $h\rightarrow 0$. Therefore, for a fixed sample size $N$, values of $h$ that are either very large or very small may lead to thresholds $ \hat{\tau}_h(N,\delta)$ that are much larger than $\sigma_{rank(T)}(T_h)$, and inequality \ref{eqnest3} implies that our procedure will underestimate $rank(T)$ for such choices of $h$. This point is illustrated by Figure \ref{fig:bw_delta_DT} $(a)$ which, for design 1 in Section \ref{secSIM}, plots the probability that $\{\hat{M}=M\}$ for different values of the bandwidth $h$, and for a sample size of $N=2000$. We leave the determination of ``good" data-driven choices of the bandwidth $h$, as well as the choice  of the kernel $K$, for future research. In our simulation studies below (Section \ref{secSIM}), we implement $\widehat{M}$   with a bandwidth $h$ given by Silverman's rule ($h\sim N^{-1/6}$ when $X\in \mathbb{R}^2$), and inequality \ref{eqnsimt} below implies that we have $\|T-\hat{T}_h\|_{op}=o_p(1)$ for this particular choice of $h$.
\end{remark}
%%\indent We now suggest an estimator of the operator $T$, which is shown to be consistent in Proposition \ref{prop1} below. A consequence of Proposition \ref{prop1}  (in conjunction with inequality \ref{eqnweyl}) is that the ordered singular values of the operator $T_h$ converge to those of the operator $T$ as $h\rightarrow 0$. Unlike Proposition \ref{prop50} above, some additional regularity conditions are needed on the density $f$ for our estimator of $T$ to be consistent; in particular, we assume in Proposition \ref{prop1} that $f$ is continuous. It is not difficult to modify the proof of the proposition to obtain a convergence rate for $\| \hat{T}-T\|$ when $f$ satisfies additional regularity conditions (twice differentiable for instance).

\begin{proposition}
\label{prop1}
Suppose that the kernel $K$ satisfies: $K \in L^1(\mathbb{R})\cap L^2(\mathbb{R})$ and $\int_{\mathbb{R}} K=1$, and let $T_h$ and $\hat{T}_{h}$ be defined as in equation \ref{eqnth} and \ref{estt}, respectively. Then we have:
\begin{equation}
\label{eqnh0}
\lim_{h\rightarrow 0} \|T_h-T\|_{HS}=0 \ \ \text{and} \ \ \ \lim_{h\rightarrow \infty} \|T_h\|_{HS}=0
\end{equation}
Moreover, if the bandwidth $h$ is chosen such that $h=h(N)\rightarrow 0$ and $Nh^2\rightarrow \infty$, as $N\rightarrow \infty$, then $\hat{T}_{h}$  is a consistent estimator of $T$:
\begin{equation}
\label{eqnsimt}
E\| \hat{T}_h-T\|_{op}\leq \sqrt{E\| \hat{T}_h-T\|_{HS}^2}=o(1).
\end{equation}

\end{proposition}
As a consequence of equation \ref{eqnh0}  (in conjunction with inequality \ref{eqnweyl}) the singular values of $T_h$ converge to those of $T$ as $h\rightarrow 0$ (in particular the smallest nonzero singular value of $T_h$ converges to the smallest nonzero singular value of $T$), and the singular values of $T_h$ all tend to zero as $h\rightarrow \infty$. Similarly, inequality \ref{eqnsimt} implies that if $h$ does not decay to zero too fast as $N \rightarrow \infty$ ($Nh^2\rightarrow \infty$), then the singular values of  $\hat{T}_{h}$ are arbitrarily close to the corresponding singular values of $T$ with probability approaching one (as $N\rightarrow \infty$). The proof of inequality \ref{eqnsimt} involves the decomposition of the error $\| \hat{T}_h-T\|_{HS}$ into an approximation bias that controls the difference $T-T_h$, and an estimation error that controls the difference $\hat{T}_h-T_h$. The condition $h=h(N)\rightarrow 0$ is needed to make the approximation bias converge to zero, and the condition $Nh^2\rightarrow \infty$ is needed to make the estimation error converge to zero. It is not difficult to modify the proof of the proposition to obtain a convergence rate for $\| \hat{T}-T\|_{op}$ when the density $f$ satisfies additional regularity conditions (twice differentiable with compact support for instance).

 Figure \ref{fig:box_whiskers} ((a) and (b)) provides an illustration of Proposition \ref{prop1}; it shows the five largest singular values of the operator $\hat{T}_h$, for the specific choice of bandwidth: $h=0.05$. For this choice of $h$, since $\hat{T}_h$ is a consistent estimator of $T_h$, we expect the singular values of $\hat{T}_h$ to be arbitrarily close (with high probability) to those of $T_h$, as  $N$ gets large.  Moreover, since $h$ is ``relatively small", we expect the singular values of $T_h$ to be close to those of $T$. Figures \ref{fig:box_whiskers} ((a) and (b)) correspond to design 2 in Section \ref{secSIM}, where the data is generated from a mixture of three uniforms with equal weights: $\pi_0=\pi_1=\pi_3=1/3$. As noted in Remark \ref{remark0} (equation \ref{example}), the nonzero singular values of the operator $T$ for this design coincide with the mixing proportions, and we have $\sigma_1(T)=\sigma_2(T)=\sigma_3(T)=1/3$.  Note that the 3 largest singular values of the estimator $\hat{T}_h$ plotted in Figure \ref{fig:box_whiskers} (Box (a) and (b)) are all close to $1/3$.

\begin{remark}
\label{rem004}
When $K=2$ and at least one of the components of $X$ has dimension greater than 1 ($d_1\neq 1$ and/or $d_2\neq 1$), the results in Proposition \ref{prop50} and Theorem \ref{theo1} remain valid with $T_h$ and $\hat{T}_h$ defined as follows: For $d \in \{d_1,d_2\}$ and $h>0$, let $K^{d}_h:\mathbb{R}^{d}\rightarrow \mathbb{R}$ be the $d$-fold product of $K_h$, i.e., $K^d_h(x)=\prod_{j=1}^d K_h(x_j)$ for all $x \in \mathbb{R}^d$. Then the operator $\hat{T}_h$ is as defined in equation \ref{estt} (with $w$ now in $L^2(\mathbb{R}^{d_1})$), with $\hat{f}_h$ now given by 
\begin{equation}
\label{eqnext1}
\hat{f}_h(x_1,x_2)=\frac{1}{N}\sum_{i=1}^N K^{d_1}_h(x_1-X^1_i)K^{d_2}_h(x_2-X^2_i),
\end{equation}
for all $x_1\in \mathbb{R}^{d_1}$ and $x_2\in \mathbb{R}^{d_2}$. Similarly, the operator $T_h$ is defined as in equation \ref{eqnth}, with the function $f_h$ now replaced by $f_h(x_1,x_2)=E[K^{d_1}_h(x_1-X^1)K^{d_2}_h(x_2-X^2)]$.

 When $K\geq 2$, then the preceding construction can be applied to any variable pair $(X^i,X^j)$ ($1\leq i<j\leq K$),
and, following Proposition \ref{prop04}, use the maximum of the estimates of the ranks of operators associated with each pair $(X^i,X^j)$ as an estimator of $M$. 
%A second approach, similar to that taken in Kasahara and Shimotsu \cite{HKKS} (see p. 101), consist in grouping the variables in $X=(X^1,\cdots,X^K)$ into two groups, and apply the results of the case $K=2$.

\end{remark}

%%%%%%%%%%%%%%%%%%%%%%%%%%%%%%%%%%%%%%%%%%%%%%%%%%%%%
%%% INSERT FIGURE SINGULAR VALUE BOX AND WHISKERS %%%
%%%%%%%%%%%%%%%%%%%%%%%%%%%%%%%%%%%%%%%%%%%%%%%%%%%%%

%%%%%%%%%%%%%%%%%%%%%%%%%%%%%%%%%%%%%%%%%%%%%%%%%%%%%%%%%%%%%%%%%%%%%%%%%%%%%%%%%%%%%%%%%%%%%%%%%%%%%%%%%%%%%%%%%%%%%%%%%%%%%%%%%%%%

%Selection frequencies of $M =3$ when varying $(h)$ or $\delta$ using Design 1 and when $N=2000$

\subsection{Computation of singular values}
\label{secIMP}
To evaluate $\widehat{M}$ in Theorem  \ref{theo1}, it is necessary to provide a procedure for computing the singular values of $\hat{T}_h$. Let $\hat{T}_h$ be as in equation \ref{estt}. We show in the proof of Corollary \ref{cor1} below that the singular values of $\hat{T}_h$ are equal to the singular values of the matrix $\hat{A}_h$ defined by :
\begin{equation}
\label{eqn01}
    \hat{A}_h=\frac{\hat{W}_{2h}^{1/2}\hat{W}_{1h}^{1/2}}{N},
\end{equation}
Here the matrices $\hat{W}_{1h}$ and $\hat{W}_{2h} \in \mathbb{R}^{N \times N}$ are given by
\begin{equation}
\label{eqn02}
    [\hat{W}_{1h}]_{i,j}=\phi_h(X^1_{i},X^1_{j})\  \ \text{and}\ \ \  [\hat{W}_{2h}]_{i,j}=\phi_h(X^2_{i},X^2_{j})
\end{equation}
for $1\leq i,j \leq N$, with the function $\phi_h$ ($\phi_h:\mathbb{R}^2\rightarrow \mathbb{R}$) defined by 
\begin{equation}
\label{eqnphi}
    \phi_h(a,b)=\int K_h(a-u)K_h(b-u) du,
\end{equation}
for all $(a,b)\in \mathbb{R}^2$. Note that the function $\phi_h$ can be computed in closed form for many choices of the kernel $K$:  for instance, $  \phi_h(a,b)=(2h\sqrt{\pi})^{-1}\exp{\left(-\frac{(a-b)^2}{4h^2}\right)}$ if the kernel $K$ is Gaussian ($K(x)=\frac{\exp{-x^2/2}}{\sqrt{2\pi}}$), and $\phi_h(a,b)=\bold{1}\{|a-b|\leq 2h\}\frac{2h-|a-b|}{4h^2}$ if the kernel $K$ is uniform ($K(x)=(1/2)\bold{1}\{|x|\leq 1\}$).
We state the foregoing observations as a corollary.
\begin{corollary}
\label{cor1}
The estimator $\widehat{M}$ of Theorem \ref{theo1} is equivalently given by  
 \begin{equation}
 \label{eq:cor1_eq}
\widehat{M}=\#\{j\ | \  r_j(\hat{A}_h)\geq \hat{\tau}_h(N,\delta)\},
\end{equation}
where the matrix $\hat{A}_h$ is as defined in equation \ref{eqn01}, and $r_j(\cdot)$ is defined as in equation \ref{eqnr}.
\end{corollary}

\begin{remark}
\label{rem0005}
As in Remark \ref{rem004}, when $K=2$ and at least one of the components of $X$ has dimension greater than 1 ($d_1\neq 1$ and/or $d_2\neq 1$), the results in Corollary \ref{cor1} remain valid with the matrix $\hat{A}_h$ defined as in \ref{eqn01}, where the matrix $\hat{W}_{1h}$ and  $\hat{W}_{2h}$ are now defined by:
\begin{equation}
\label{eqnext2}
    [\hat{W}_{1h}]_{i,j}=\phi^{d_1}_h(X^1_{i},X^1_{j}) \  \text{and} \  [\hat{W}_{2h}]_{i,j}=\phi^{d_2}_h(X^2_{i},X^2_{j})
\end{equation}
with $\phi^{d}_h:\mathbb{R}^d\times \mathbb{R}^d\rightarrow \mathbb{R}$ ($d\in \{d_1,d_2\}$) defined by
\begin{equation}
\label{eqnext3}
\phi^{d}_h(a,b)=\int_{\mathbb{R}^{d}} K^{d}_h(a-u)K^{d}_h(b-u)du.
\end{equation}

Here $K^d_h(\cdot)$ is as defined in Remark \ref{rem004}. When $K$ is the Gaussian kernel, a simple computation yields $\phi^{d}_h(a,b)=(2h\sqrt{\pi})^{-d}\exp\left(\frac{-\|a-b\|^2}{4h^2}\right)$.
\end{remark}

%%%%%%%%%%%%%%%%%%%%%%%%%%%%%%%%%%%%%%%%%%%%%%%%%%%%%%%%%%%%%%%%%%%%%%%%%%%%%%%%%%%%%%%%%%%%%%%%%%%%%%%%%%%%%%%%%%%%%%%%%%%%%%%%%%%%

\subsection{Computation of the threshold rule}
\label{secH}

In this section, we provide a numerical procedure to compute the threshold  $\hat{\tau}_h(N,\delta)$. Although the threshold $\hat{\tau}$ defined by equation \ref{eqn53} can be used to estimate $M$, we have observed in our simulations studies that this threshold produces very conservative estimates ($\widehat{M}<M$ with high probability) for small sample sizes ($N$ less than 500), thus suggesting that it may be too loose an upper bound on the estimation error. We use instead (and recommend) the threshold suggested by inequality \ref{eqntau}, which leads to a better and more reasonable performance for small sample sizes; i.e:
% threshold is the one used to implement our method in Section \ref{secSIM}, is the one suggested by inequality
for $0<\delta<1$ given, the new threshold $\hat{\tau}(N,\delta)$ is given by the value of $\tau>0$ that solves  the following equation:
\begin{equation}
\label{threshold}
\ln(\delta/2)=-\frac{\tau N}{\hat{L}_h}  \ln\left(1+\frac{\tau \hat{L}_h}{\hat{\sigma}_h^2}\right)+N\times \ln 
  \left(1+\frac{\tau}{\hat{L}_h}-\frac{\hat{\sigma}_h^2}{\hat{L}_h^2}\ln \left(1+\frac{\tau \hat{L}_h}{\hat{\sigma}_h^2}\right) \right).
\end{equation}
Here $\hat{L}_h$ and $\hat{\sigma}_h^2$ are the sample analogues of $L_h$ and $\sigma_h^2$ respectively, i.e, $\hat{L}_h=:\sup_{i\neq j} \|T_{h,X_i}-T_{h,X_j}\|_{HS} $ and $\hat{\sigma}_h^2:= \frac{1}{2N(N-1)}\sum_{i\neq j} \|T_{h,X_i}-T_{h,X_j}\|_{HS}^2$. The main drawback of this new threshold, compared to the one given by \ref{eqn53}, is that it is not available in closed form and has to be solved for numerically. For a justification of this approach, note that a slight modification of the proof of Proposition \ref{prop50} (replacing $\sigma_h^2$ in inequality \ref{eqnpin2} by 1/2 times the right-hand side of inequality \ref{eqnhoeff}) implies that for all $0<\delta<1$, the following inequality holds with probability at least $1-\delta$:
\begin{equation}
\label{threshold1}
\begin{split}
&P(\|\hat{T}_h-T_h\|_{HS}>\tau)\leq 2 \exp\left\{-\frac{\tau N}{L_h}  \ln\left(1+\frac{\tau L_h}{\hat{\Sigma}_h^2}\right)\right\} \times \\
&  \left(1+\frac{\tau}{L_h}-\frac{\hat{\Sigma}_h^2}{L_h^2}\ln \left(1+\frac{\tau L_h}{\hat{\Sigma}_h^2}\right) \right)^N \ \text{for all} \ \ \tau>0
\end{split}
\end{equation}
where $\hat{\Sigma}_h^2:=\hat{\sigma}_h^2+\frac{\hat{L}_h ^2}{2}\sqrt{\left(\frac{\ln(1/\delta)}{N}\right)}$. Hence if $\tilde{\tau}(N,\delta)$ is the value of $\tau$ such that the right-hand side of inequality \ref{threshold1} is equal to $\delta$, then $P(\|\hat{T}_h-T_h\|>\tilde{\tau}(N,\delta))\leq 2 \delta$. The right-hand side of equation \ref{threshold} (exponentiated) is obtained by droping the lower order term $\frac{\hat{L}_h ^2}{2}\sqrt{\left(\frac{\ln(1/\delta)}{N}\right)}$ from $\hat{\Sigma}_h^2$. And although the latter change is not justified by our results, it has no (relevant) effect on the performance of our procedure for large $N$. \\

 \indent To solve for $\tau$ in equation \ref{threshold}, it suffices to provide a procedure to compute   $\|T_{h,x}-T_{h,x'}\|_{HS}^2$ for any $x=(x_1,x_2)$ and $x'=(x_1',x_2')$ in $\mathbb{R}^2$. 
%(note that $\hat{L}_h= \sqrt {\sup_{i\neq j} \|T_{h,X_i}-T_{h,X_j}\|_{HS}^2} $).
 As the Hilbert-Schmidt norm is an inner product norm, we have
\begin{equation*}
\|T_{h,x}-T_{h,x'}\|_{HS}^2=\|T_{h,x}\|_{HS}^2+\|T_{h,x'}\|_{HS}^2-2\langle T_{h,x}, T_{h,x'}\rangle_{HS},
\end{equation*}
where $\langle \cdot,\cdot \rangle_{HS}$ denotes the Hilbert-Schmidt inner product. A straightforward computation (using the definition of the Hilbert-Schmidt inner product) yields 
 \begin{equation}
 \label{eqn00001}
 \begin{aligned}
 &\|T_{h,x}-T_{h,x'}\|_{HS}^2\\
 &=\phi_h(x_1,x_1)\phi_h(x_2,x_2)+\phi_h(x_1',x_1')\phi_h(x_2',x_2')-2\phi_h(x_1,x_1')\phi_h(x_2,x_2'),
 \end{aligned}
 \end{equation}
with the function $\phi_h$ defined by equation \ref{eqnphi}.
\begin{remark}
\label{rem00006}
As in Remark \ref{rem004} and \ref{rem0005}, when $K=2$ and at least one of the components of $X$ has dimension greater than 1 ($d_1\neq 1$ and/or $d_2\neq 1$), the threshold rule given by equation \ref{threshold} remains valid, and the only modification is that we need to replace the functions $\phi_h$ in the right-hand side of equation \ref{eqn00001} by the functions $\phi^d_h$ (equation \ref{eqnext3}), in order to compute the quantities $\|T_{h,x}-T_{h,x'}\|_{HS}^2$.
\end{remark}
%%%%%%%%%%%%%%%%%%%%%%%%%%%%%%%%%%%%%%%%%%%%%%%%%%%%%%%%%%%%%%%%%%%%%%%%%%%%%%%%%%%%%%%%%%%%%%%%%%%%%%%%%
%%%%%%%%%%%%%%%%%%%%%%%%%%%%%%%%%%%%%%%%%%%%%%%%%%%%%%%%%%%%%%%%%%%%%%%%%%%%%%%%%%%%%%%%%%%%%%%%%%%%%%%%%
%%%%%%%%%%%%%%%%%%%%%%%%%%%%%%%%%%%%%%%%%%%%%%%%%%%%%%%%%%%%%%%%%%%%%%%%%%%%%%%%%%%%%%%%%%%%%%%%%%%%%%%%%

\section{Monte Carlo Experiments}
\label{secSIM}

In this section, we assess the performance of our estimator $\widehat{M}$ on five designs. The performance of $\widehat{M}$ is then compared to the  procedures suggested by Kasahara and Shimotsu \cite{HKKS}: SHT, AIC, BIC, HQ (when $K=2$) and $\text{max-rk}^{+}$ (when $K>2$). The designs that we consider have $M=3$ and $M=5$ mixture components, and for each design we  simulate 1000 samples of sizes $N=500$ and $N=2000$.  The first four designs are bi-variate ($K=2$), but in order to assess the performance of our estimator when more than two conditionally independent variables are observed, we include a design with $K=8$.

To compute $\widehat{M}$ for each Monte Carlo sample, we construct the matrix $\hat{A}_h$ defined in Equation \ref{eqn01}, and compute its singular values. We use the Gaussian kernel, i.e. $K(x)=\frac{\exp{-x^2/2}}{\sqrt{2\pi}}$, and the bandwidth $h$ is chosen according to Silverman's rule. Finally, we use the threshold rule $\hat{\tau}_h(N,\delta)$ given by equation \ref{threshold}, with $\delta=0.05$ for all of our simulations. For designs with $K>2$ variables (Design 5), we compute $\widehat{M}$  for each of the $K \choose 2$ pairs of variables, and use the maximal values of all such estimates as our estimate of $M$ (as suggested by Proposition \ref{prop04}).

We consider the following five designs for our simulations. Designs 1 and 5 are from Kasahara and Shimotsu \cite{HKKS}, and Designs 2, 3 and 4  highlight different aspects of the data generating process that affect the performance of our procedure.  \ 

\begin{enumerate}

\item Design 1  (mixture of 3 normal distributions): \\* 
$P(\Theta=m)=1/3$ for $m\in \{1,2,3\}$, and $(X^1,X^2)|\Theta=m\sim \mathcal{N}(\mu_m,I_{2})$,
where $\mu_{1}=(0,0)',$ $\mu_{2}=(1,2)'$, $\mu_{3}=(2,1)'$, and $I_{2}$ is the 2 by 2 identity matrix.

\item Design 2 (mixture of 3 Uniform distributions): \\*
$P(\Theta=m)=1/3$ for $m\in \{1,2,3\}$, and $(X^1,X^2)|\Theta=m\sim \mathcal{U}(a_m,b_m)\times \mathcal{U}(a_m,b_m)$, with
$a_m=(m-1)$ and $b_m=m$.

\item Design 3 (mixture of 3 normal distributions): \\*
$P(\Theta=m)=1/3$ for $m\in \{1,2,3\}$, and $(X^1,X^2)|\Theta=m\sim \mathcal{N}(\mu_{m},I_{2})$,
where $\mu_{1}=(0,0)',$ $\mu_{2}=(3,3)'$, $\mu_{3}=(-3,-3)'$, and $I_{2}$ is the 2 by 2 identity matrix.

\item Design 4 (mixture of 5 uniform distributions):\\* 
$P(\Theta=m)=1/5$ for $m\in \{1,2,3,4,5\}$, and $(X^1,X^2)|\Theta=m\sim \mathcal{U}(a_m,b_m)\times \mathcal{U}(a_m,b_m)$, with
$a_m=(m-1)$ and $b_m=m$.

\item Design 5 (mixture of 3 normal distributions):
$P(\Theta=m)=1/3$ for $m\in \{1,2,3\}$, and $(X^1,X^2,\cdots,X^8)|\Theta=m \sim \mathcal{N}(\mu_{m},I_{8})$, where
with $\mu_{1}=(0,0,0,0,0,0,0,0)', \mu_{2}=(1.0,2.0,0.5,1.0,0.75,1.25,0.25,0.5)', 
\mu_{3}=(2.0,1.0,1.0,0.5,1.25,0.75,0.5,0.25)'$, and $I_{8}$ is the 8 by 8 identity matrix.

\end{enumerate}

The outcome of the simulations are presented in the tables below (one table for each design). The implementation of the method of Kasahara and Shimotsu \cite{HKKS} requires us to choose a value for the parameter $M_0$. We recall that the parameter $M_0$ in their procedure represents a guess by the analyst of an upper bound on $M$, and they recommend using a partition $\Delta=\Delta^1\times \Delta^2$ of size $M_0^2$ ($|\Delta^1|=|\Delta^2|=M_0$) when implementing their procedures. When we implement their procedures, for design 1 through 4, we consider the choices $M_0=4$ and $M_0=8$. The partitions $\Delta$ are then constructed by partitioning the supports of $X^1$ and $X^2$ into $M_0$ equiprobable (with respect to the empirical distribution) intervals as suggested by Kasahara and Shimotsu \cite{HKKS}.
We implement Design 5 exactly as in Kashara and Shimotsu \cite{HKKS}, and use their max $\text{ave-rk}^{+ }$ statistic, with $M_0=4$, to estimate $M$ (see p. $107-108$ of \cite{HKKS} for details).

%%%%%%%%%%%%%%%%%%%%%%%%%%%%%
%%  TABLES %%
%%%%%%%%%%%%%%%%%%%%%%%%%%%%%

%\newgeometry{left=3cm,bottom=0.1cm}

\begin{center}
\begin{adjustbox}{max width = \linewidth}
\begin{threeparttable}[H]
\caption{Design 1}
%\vspace{-1cm}
\input{Tables/design1.tex}
\end{threeparttable}
\end{adjustbox}
\end{center}

\begin{center}
\begin{adjustbox}{max width = \linewidth}
\begin{threeparttable}[H]
\caption{Design 2}
%\vspace{-1cm}
\input{Tables/design2.tex}

\end{threeparttable}
\end{adjustbox}
\end{center}

\begin{center}
\begin{adjustbox}{max width = \linewidth}
\begin{threeparttable}[H]
\caption{Design 3}
%\vspace{-1cm}
\input{Tables/design3.tex}

\end{threeparttable}
\end{adjustbox}
\end{center}

\begin{center}
\begin{adjustbox}{max width = \linewidth}
\begin{threeparttable}[H]
\caption{Design 4}
%\vspace{-1cm}
\input{Tables/design4.tex}

\end{threeparttable}
\end{adjustbox}
\end{center}

\begin{center}
\begin{adjustbox}{max width = \linewidth}
\begin{threeparttable}[H]
\caption{Design 5}
%\vspace{-1cm}
\input{Tables/design6.tex}

\end{threeparttable}
\end{adjustbox}
\end{center}

\indent Our method performs the worst in Design 1 (relative to the other designs), and selects the true number of components only $40\%$ of the time when $N=2000$. As noted in Remark \ref{rem001}, we expect our approach to yield conservative estimates of $M$ if the smallest non-zero singular value of the operator $T$ is close to zero (relative to the sample size). From Figure \ref{fig:box_whiskers} (Box (d)), we see that the (estimated) third largest singular value of $T$ in Design 1 is very small (approximately equal to $0.02$). Note however that the performance of our procedure is comparable to the second best procedure of \cite{HKKS} (SHT), and their best procedure AIC selects the correct number of components $61\%$ of the time when $N=2000$.\\
\indent In Design 2, all nonzero singular values of $T$ are equal to $1/3$ (see Remark \ref{rem00}), hence much larger in magnitude than those of design 1. And as can be expected from inequality \ref{eqnest2}, our estimator performs quite well; $\widehat{M}$ always selects 3 components  for both sample sizes. By contrast, all the methods of Kasahara and Shimotsu \cite{HKKS} perform poorly on this design, with their best method (BIC) selecting $M=3$ with a frequency  of approximately $50\%$ when $N=2000$ and $M_0=4$. Moreover, all of their estimation procedures tend to substantially overestimate the true number of components when $M_0=8$, with AIC selecting $M\geq 4$ approximately $30\%$ of the time when $N=2000$. From this design and Design 4 below, we observe that the methods of Kasahara and Shimotsu \cite{HKKS} seem to perform poorly when the support $X$ is ``irregular" and the matrix $P_{\Delta}$ is sparse (has many zeros). \\
\indent Design 3 combines the desirable aspects of Design 1 and 2: the variable $X$ has full support as in Design 1, and the nonzero singular values of the operator $T$ have moderate size as in Design 2 (from simulations $\sigma_3(T)\approx 0.1$). Our method as well as the procedures of Kasahara and Shimotsu \cite{HKKS} perform well on this design. However, the performance of their procedures decrease when the number of partitions is increased ($M_0=8$), and AIC tends to overestimates the number $M$ even when $N=2000$ (by as much as $20\%$ of the time when $M_0=8$). As noted in Kasahara and Shimotsu \cite{HKKS}, the method AIC is not necessarily consistent, and it tends to overestimate the rank of $P_{\Delta}$ when $N$ is large. \\
\indent Design 4 is a variation of Design 2 (also a mixture of uniforms), where $M=5$ and the nonzero singular values of $T$ are smaller (all five nonzero singular values of $T$ are equal to $1/5$). As the nonzero singular values of $T$ are smaller in comparison to those of design 2, the performance of our method deteriorates relative to design 2. Indeed, our method now selects the true number of components approximately $95\%$ of the time when $N=500$. However, when $N=2000$ our method selects the true number of components in all of the Monte Carlo samples. As in Design 2, the methods of Kasahara and Shimotsu \cite{HKKS} do not perform well on this design. We recall here that given an upper bound $M_0$ on $M$, the procedures of  Kasahara and Shimotsu \cite{HKKS} yield an estimate of a lower bound on $M$ that is at most equal to $M_0$. We see here that when the upper bound is incorrectly set at $M_0=4$, all of their procedures select $M=4$ approximately $50\%$ of the time when $N=2000$. When $M_0=8$, all of their procedures select the true number of components in approximately $10\%$ of the simulations when $N=2000$. As noted above, the poor performance of their procedures is probably due to the fact that the support of $X$ is highly ``irregular" and that the matrices $P_{\Delta}$ are sparse.

Finally, Designs 5 assesses the performance of our estimator when $K>2$ (here $K=8$). For $N=500$,  our procedure performs poorly and only selects the correct number of components approximately  $1\%$ of the time.  This performance is comparable to the  BIC (ave-rk) and HQ (ave-rk) procedures  of Kashara and Shimotsu \cite{HKKS}, which selects the correct number of components $0.01\%$ and $0.013\%$ of the time, respectively. But our performance here is significantly  worse than their AIC (ave-risk) procedure which selects the correct number of components $11.9\%$ of the time. However, when the sample size is increased to $N=2000$, our method performs significantly better and correctly chooses the number of components $50.7\%$ of time. In comparison, none of Kashara and Shimotu's \cite{HKKS} procedures  enjoy the same jump in performance: their best performer (AIC\ ave-rk) is correct 39.9\% of the time while their worst performer (BIC  ave-rk) is correct only $0.023\%$  of the time. \\
\indent Note that across all designs, the probability of overestimating $M$ is much smaller than the upper bound of $2\delta(=0.1)$ in Design 1 through 4, and the (crude) upper bound of $ {K \choose 2} 2 \delta $ in Design 5; in fact, in all our simulations, our method never overestimates $M$ (by contrast, at  least three of the procedures of Kashara and Shimotu \cite{HKKS} overestimate the number of components in all designs).  In Table \ref{tabledelta}, we assess the performance of our method on Design 1 for various values of $\delta$. Note that for all values of $\delta$ considered,  $P(\widehat{M}>M)$ is much smaller than the upper bound of $2\delta$ suggested by inequality \ref{eqnest1} (in fact when $\delta=0.5$ we overestimate $M$ less than 1\% of the time when $N=2000$, and never overestimate $M$ when $N=500$). This potential slackness in inequality \ref{eqnest1} suggests that it might be desirable to use larger values of $\delta$. In Table \ref{alldesigns}, we provide the simulations outcomes when our method is reapplied to all 5 designs with $\delta=0.4$. We see a general improvement in performance across all designs (at the exception of Design 2), we now select the correct number of components at a higher frequency, and the probability of overestimating $M$ is less than 3\% in all designs.

\begin{center}
\begin{adjustbox}{max width = \linewidth}
\begin{threeparttable}[H]
\caption{Design 1, variable $\delta$}
%\vspace{-1cm}
\input{Tables/design8.tex}
\label{tabledelta}
\end{threeparttable}
\end{adjustbox}
\end{center}

\begin{center}
\begin{adjustbox}{max width = \linewidth}
\begin{threeparttable}[H]
\caption{All Designs with $\delta = 0.40$ }
%\vspace{-1cm}
\input{Tables/delta.tex}
\label{alldesigns}
\end{threeparttable}
\end{adjustbox}
\end{center}

%%%%%%%%%%%%%%%%%%%%%%%%%%%%%%%%%%%%%%%%%%%%%%%%%%%%%%%%%%%%%%%%%%%%%%%%%%%%%%%%%%%%%%%%%%%%%%%%%%%%%%%%%%%%%%%%%%%%%%%%%%%%%%%%%%%%%%%%%%%%%%%%%%%%%%
%%%%%%%%%%%%%%%%%%%%%%%%%%%%%%%%%%%%%%%%%%%%%%%%%%%%%%%%%%%%%%%%%%%%%%%%%%%
%%%%%%%%%%%%%%%%%%%%%%%%%%%%%%%%%%%%%%%%%%%%%%%%%%%%%%%%%%%%%%%%%%%%%%%%%%%%%%%%%%%%%%%%%%%%%%%%%%%%%%%%%%%%%%%%%%%%%%%%%%%%%%%%%%%%%%%%%%%%%%%%%%%%%%
%%%%%%%%%%%%%%%%%%%%%%%%%%%%%%%%%%%%%%%%%%%%%%%%%%%%%%%%%%%%%%%%%%%%%%%%%%%

\section{Empirical Examples}
\label{secA}
\begin{figure}[H]
{\includegraphics[width=0.49\textwidth]{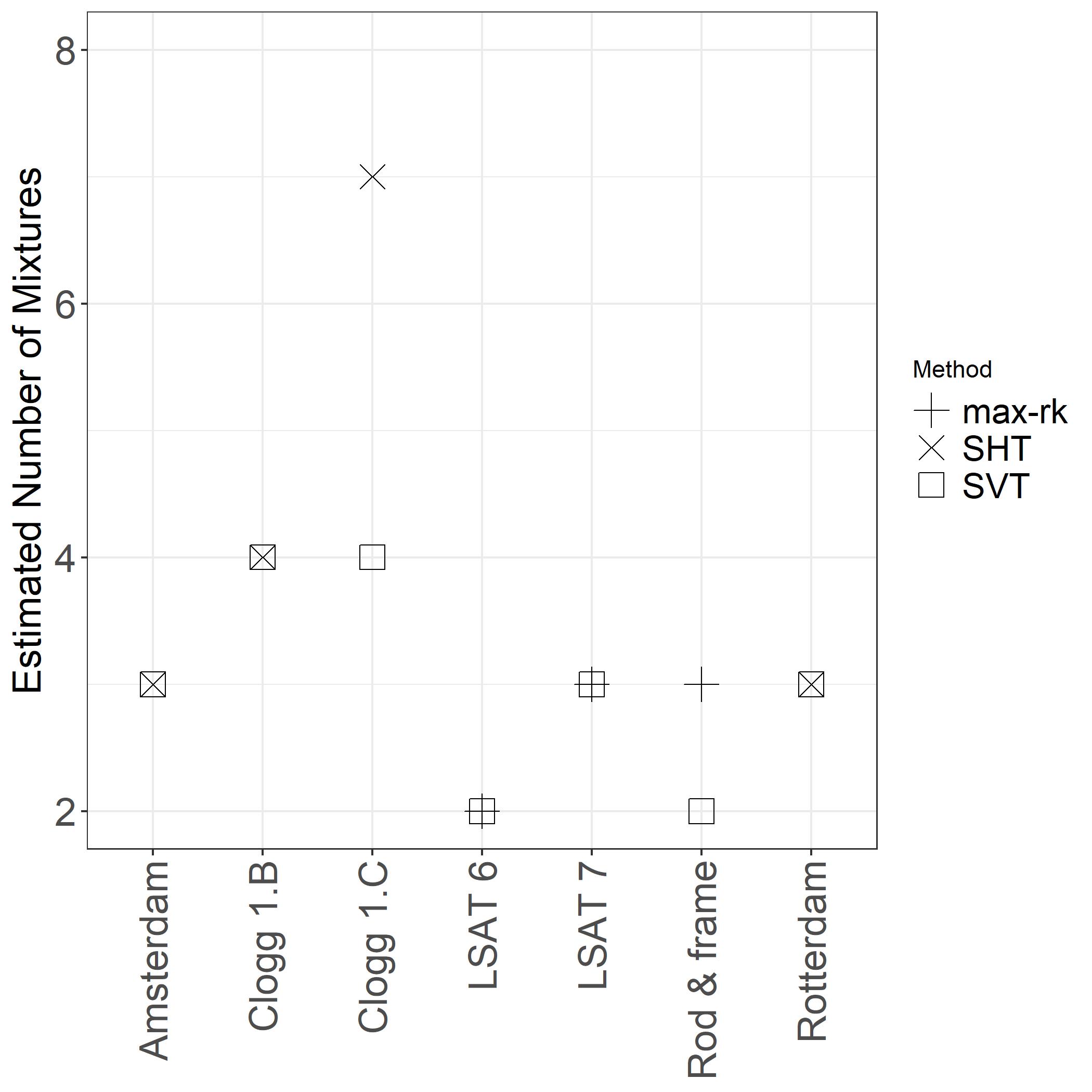}}
\hfill % <-- Seperation
{\includegraphics[width=0.49\textwidth]{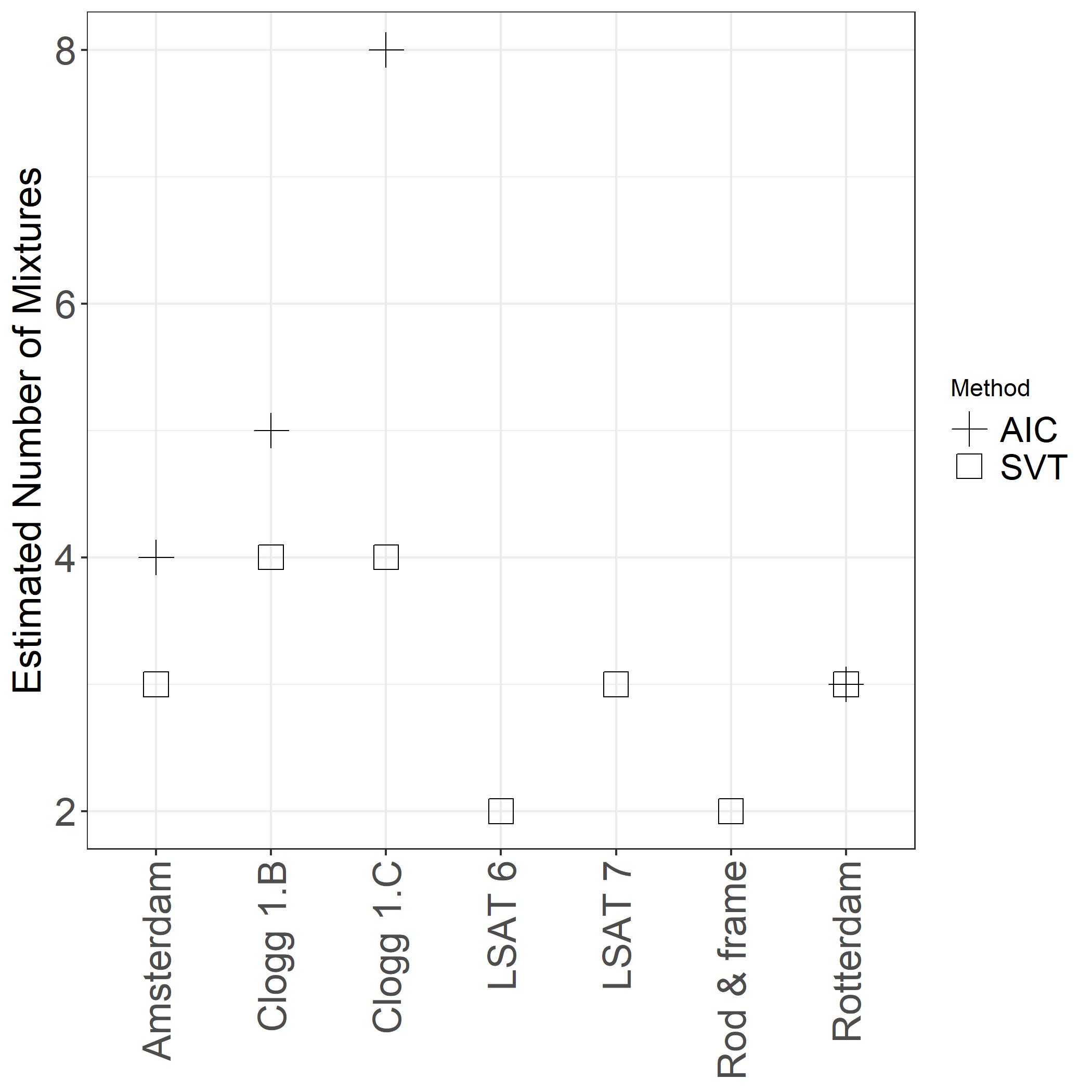}}
\\ % <-- Line break
{\includegraphics[width=0.49\textwidth]{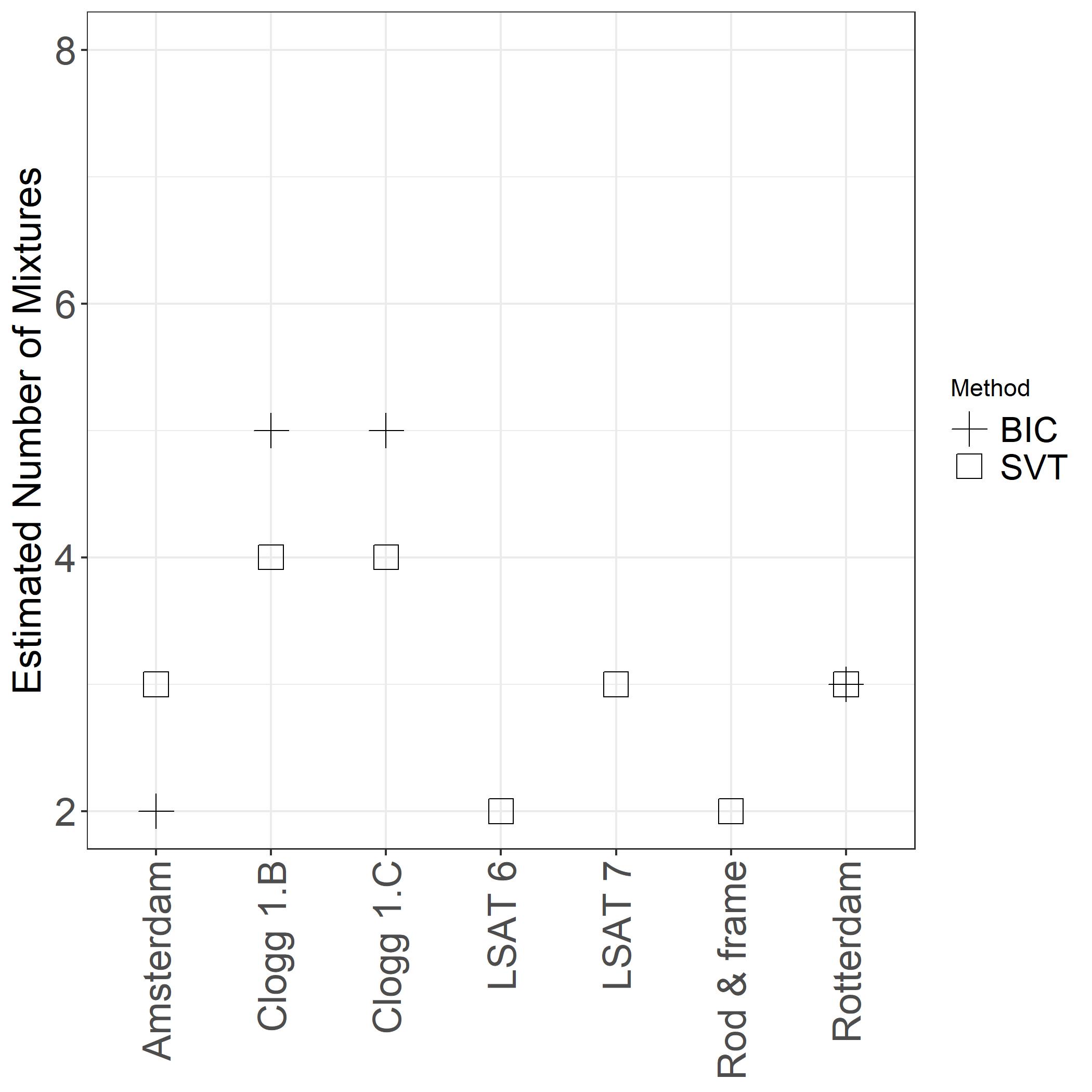}}
\hfill % <-- Seperation
{\includegraphics[width=0.49\textwidth]{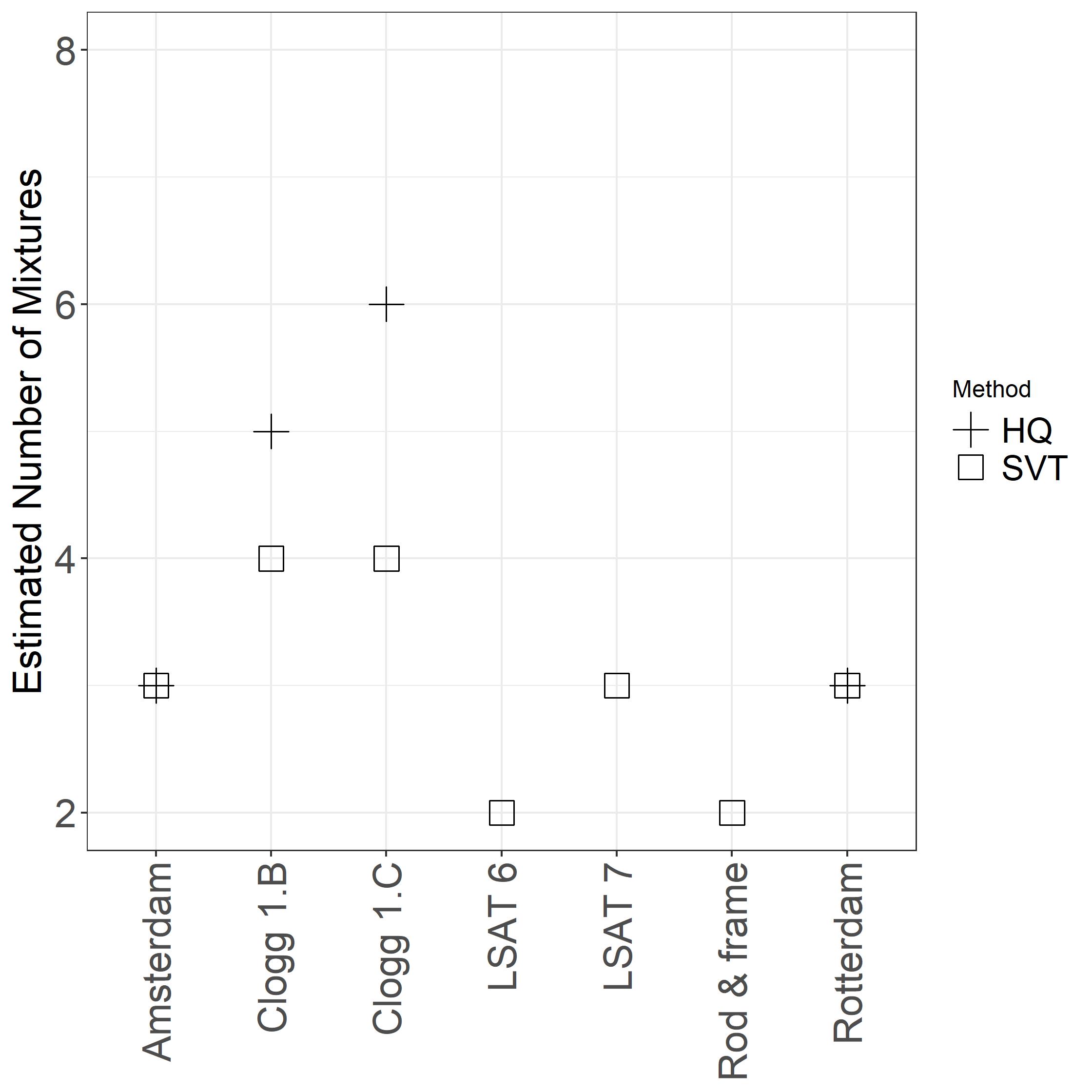}} 
\caption{Estimates for each empirical example compared with estimates obtained using the procedures suggested by  Kashara and Shimotsu \cite{HKKS}. \ }
\label{fig:emp_DT}
\end{figure}

In this section, following Kasahara and Shimotsu \cite{HKKS}, we  apply our estimator $\widehat{M}$ on four empirical examples  containing seven datasets. These datasets are obtained from Clogg \cite{clogg}, Van der Heijden et al. \cite{vdv}, Mislevy \cite{mislevy} and Hettmansperger and Thomas \cite{hettman}. We  provide below only a brief description of each dataset,  and refer the interested reader to the original papers for a more detailed discussion of each empirical example.\\
\indent As in our Monte Carlo simulations (Section \ref{secSIM}), we implement our estimator $\widehat{M}$ using: the Gaussian kernel; the bandwidth $h$  chosen according to Silverman's rule; the threshold rule $\hat{\tau}_h(N,\delta)$ given by equation \ref{threshold}; and  the parameter $\delta =0.05$ (we also provide for each dataset the estimated number of mixture components when $\delta=0.4$). For empirical examples where there are more than two conditionally independent variables (at the exception of the LSAT datasets), we compute $\widehat{M}$ for each  of the $K \choose 2$ pairs of variables, and use the maximum of these estimates as our estimate of $M$, as suggested by Proposition \ref{prop04}. For the LSAT datasets, we take the approach suggested by Proposition \ref{prop005}.\\
\indent In Figure \ref{fig:emp_DT}, we graphically display our estimates alongside the estimates obtained by Kasahara and Shimotsu \cite{HKKS} (see p. 108-110 of \cite{HKKS}, for a detailed discussion of the implementation of their procedures for these empirical examples). When $K=2$ we compare our procedure to each of the estimates obtained by the four procedures suggested by Kasahara and Shimotsu \cite{HKKS}: SHT (with the parameter $\alpha=0.05$), AIC, BIC and HQ; when $K>2$, we compare our procedure to the estimate given by the  $\text{max-rk}^{+}$ statistic (with the parameter $\alpha=0.05$) of  \cite{HKKS}.\\
\indent As a general remark, our estimator $\widehat{M}$ agrees with at least one of their procedures in five of the seven datasets, and in the other two datasets their estimates are larger than ours. Furthermore, at the exception of the two datasets from Clogg \cite{clogg} our estimates are the same for $\delta = 0.05$ and $\delta = 0.40$ (when $\delta=0.4$ $\widehat{M}=5$ for the two datasets of Clogg \cite{clogg}).

\subsection{Intergenerational occupational mobility in Britain}
In the first empirical example, we estimate the number of mixture components in two datasets describing intergenerational occupational mobility studied by Clogg \cite{clogg}. The first dataset (Table 1.C of Clogg \cite{clogg}) contains 3,497 pairs of father-son occupations $(K=2)$. Occupations are grouped into a total of 8 categories: (1) professional
and high administrative; (2) managerial and executive; (3) inspectional, supervisory, and other non-manual (high grade);
(4) inspectional, supervisory, and other non-manual (low grade); (5) routine grades of nonmanual; (6) skilled manual; (7)
semi-skilled manual; and (8) unskilled manual. The second dataset (Table 1.B of Clogg \cite{clogg})  is a less granular version  of the first dataset where  categories 2 and 3, and categories 6 and 7 are merged.  

\indent Applying our procedure, we estimate $\widehat{M}=4$ for both datasets (and both estimates are equal to 5 when $\delta=0.4$). By contrast, for the first dataset, Kashara and Shimotsu's \cite{HKKS} estimate 7, 8, 5 and 6  mixture components using SHT, AIC, BIC and HQ, respectively. And for the second dataset the SHT procedure estimates 4 components, while AIC, BIC and HQ estimate 5 components.

\subsection{Ethnic groups and types of trade}
 In the second empirical example, we estimate the number of mixture components in two datasets describing ethnic groups and their trading behavior in Amsterdam and Rotterdam, studied by Van der Heijden et al. \cite{vdv}. The underlying hypothesis here is that unobserved characteristics  of ethnic groups (e.g., mastery of the Dutch language) may explain their trading behavior. The datasets for  Amsterdam and Rotterdam contain  2,422 and 1,682 pairs respectively of different ethnic groups and the types of trade that they engage in $(K=2)$. 

We estimate $\widehat{M}=3$ mixture components in both Amsterdam and Rotterdam. These estimates are consistent with Van der Heijiden et al \cite{vdv} who conjectured that 3 mixture components is ``adequate''. By contrast, for the Amsterdam dataset, Kashara and Shimotsu \cite{HKKS} estimate 3, 4, 2 and 3 mixture components, using SHT, AIC, BIC and HQ, respectively. For the Rotterdam dataset, like our method, all their procedures estimate 3 mixture components.  

\subsection{Response patterns in the LSAT}
  Mislevy \cite{mislevy} studies the response patterns from  of 1000 subjects to two  subsets of the Law School Admissions Test (LSAT-6 and LSAT-7). For both surveys, each response contains 1000 responses, where each response contains five binary variables $(K=5$). \\
\indent Since we have $K>2$ binary variables, as suggested by Proposition \ref{prop005}, we consider partitions of the five variables into two groups, estimate the ranks of the operators associated with each partition, and use the maximum of the estimated ranks as our estimate of $M$. This yields an estimate of estimates 2 components for the LSAT-6 dataset, and 3 components for the LSAT-7 dataset. These two estimates are equal to those obtained by Kasahara and Shimotsu \cite{HKKS} using their $\text{max-rk}^{+}$ procedure.\\
\indent Recall that (see the discussion preceding Proposition \ref{prop005}) since the components of $X$ are binary, applying the procedure suggested by Proposition \ref{prop04} (compute $\widehat{M}$ for each of the $K \choose 2$ variable pairs, and use the maximum as an estimate of $M$) yields an estimate at most equal to 2, and we indeed get an estimate of 2 (for both datasets) when we take that approach.

\subsection{Witkin's rod and frame task } 
In our last empirical example, we estimate $M$ in a dataset that contains 83 observations of eight replications of Witkin's rod-and-frame task $(K=8)$. Each replication is measured as the rod's error deviation in degrees from the vertical. Unlike Hettmansperger and Thomas \cite{hettman}, we use both the original data and the binarized data (\cite{hettman} transform each response variable to equal 1 if its absolute value is less than or equal to 5, and 0 otherwise). For the original data, we estimate $M$ using the procedure suggested by Proposition \ref{prop04}, and for the binarized data, we use the procedure suggested by Proposition \ref{prop005}.\\
\indent Both approaches estimate 2 mixture components, while  Kashara and Shimotsu \cite{HKKS} estimate 3 components using their $\text{max-rk}^{+}$ statistic, and both estimates are comparable to those of Hettmansperger and Thomas \cite{hettman}, who obtain estimates of $M= 4, M=2$, and $M=3$, using Lindsay's gradient function method, Hellinger and Pearson's penalized distances, and the bootstrapped likelihood ratio test, respectively.

%%%%%%%%%%%%%%%%%%%%%%%%%%%%%%%%%%%%%%%%%%%%%%%%%%%%%
%%% INSERT FIGURE SINGULAR VALUE BOX AND WHISKERS %%%
%%%%%%%%%%%%%%%%%%%%%%%%%%%%%%%%%%%%%%%%%%%%%%%%%%%%%

%%%%%%%%%%%%%%%%%%%%%%%%%%%%%%%%%%%%%%%%%%%%%%%%%%%%%%%%%%%%%%%%%%%%%%%%%%%%%%%%%%%%%%%%%%%%%%%%%%%%%%%%%
%%%%%%%%%%%%%%%%%%%%%%%%%%%%%%%%%%%%%%%%%%%%%%%%%%%%%%%%%%%%%%%%%%%%%%%%%%%%%%%%%%%%%%%%%%%%%%%%%%%%%%%%%
%%%%%%%%%%%%%%%%%%%%%%%%%%%%%%%%%%%%%%%%%%%%%%%%%%%%%%%%%%%%%%%%%%%%%%%%%%%%%%%%%%%%%%%%%%%%%%%%%%%%%%%%%
\section{Conclusion}
\label{secC}
In this paper, we introduced a novel approach for estimating the number of mixture components in multivariate finite mixture models. Under a mild assumption on the distributions of the observed variables, we showed that the number of mixture components $M$ is identified and equal to the rank of an integral operator $T$, which is identified from the data. The estimator of $M$ that we proposed essentially counts the number of singular values of an estimate of a regularized version of $T$ above a data-driven threshold. We showed that our estimator is consistent, and provided finite sample performance guarantees. We presented simulation studies which showed that our estimator performs well for samples of moderate size, and we also applied our procedure to four empirical examples. Despite our method being consistent, we do not have a theory to guide the choice of many of the parameters that are involved in the estimation procedure. Looking forward, it would be interesting to address the questions of how to choose the kernel $K$ and the bandwidth $h$ in an optimal way. We leave such interesting questions for further research. 

%%%%%%%%%%%%%%%%%%%%%%%%%%%%%%%%%%%%%%%%%%%%%%%%%%%%%%%%%%%%%%%%%%%%%%%%%%%%%%%%%%%%%%%%%%%%%%%%%%%%%%%%%
%%%%%%%%%%%%%%%%%%%%%%%%%%%%%%%%%%%%%%%%%%%%%%%%%%%%%%%%%%%%%%%%%%%%%%%%%%%%%%%%%%%%%%%%%%%%%%%%%%%%%%%%%
%%%%%%%%%%%%%%%%%%%%%%%%%%%%%%%%%%%%%%%%%%%%%%%%%%%%%%%%%%%%%%%%%%%%%%%%%%%%%%%%%%%%%%%%%%%%%%%%%%%%%%%%%
\section{Proofs}
\label{appendix}

\begin{proof}{\bf (Proof of Proposition \ref{prop01})}
By equation \ref{eqnmix} $T$ has the representation $T=\sum_{m=1}^M \pi_m \ f^2_m\otimes f^1_m$. Let ${\cal M}_1$ (resp. ${\cal M}_2$) denote the subspace of $L^2({\cal S}_1)$ (resp $L^2({\cal S}_2)$) spanned by the functions $\{f^1_m\}_{m=1}^M$ (resp. $\{f^2_m\}_{m=1}^M$ ). Under Assumption \ref{FR}, the subspaces ${\cal M}_1$ and ${\cal M}_2$ have dimension $M$. Let $\langle\cdot,\cdot \rangle_1$ denote the inner product on $L^2({\cal S}_1)$. For  $\omega \in L^2({\cal S}_1)$, we have
\begin{equation*}
T(\omega)=\sum_{m=1}^M \pi_m \  f^2_m  \langle f^1_m,\omega \rangle_1
\end{equation*}
which is an element of ${\cal M}_2$, and the range of the operator $T$ is thus a subspace of ${\cal M}_2$ which has dimension at most $M$ (dimension equal to $M$ when \ref{FR} holds). To show that the range of $T$ has dimension $M$ under Assumption \ref{FR}, it thus suffices to show that each $f^2_m$ belongs to the range of $T$. Let $u_m$ be equal to the residual of the projection of $f^1_m$ on the subspace of ${\cal M}_1$ spanned by the functions $\{f^1_{m'}\}_{m'\neq m}$, and define $\omega_m=u_m/\|u_m\|^2$. The latter operation is well defined by the linear independence of the functions $\{f^1_m\}_{m=1}^M$. Then $<\omega_m, f^1_{m'}>_1=\delta_{mm'}$ (the Kronecker delta), and we have $T(\omega_m)=f^2_m$. We thus conclude that range of $T$ spans ${\cal M}_2$ and it has dimension $M$.
\end{proof}
%%%%%%%%%%%%%%%%%%%%%%%%%%%%%%%%%%%%%%%%%%%%%%%%%%%%%%%%%%%%%%%%%%%%%%%%%%%%%%%%%%%%%%%%%%%%%%%%%%%%%%%%%%%%%%%%%%%%%%%%%%%%%%%%%%%%
\begin{proof}{\bf (Proof of Proposition \ref{prop02})}
From equations \ref{eqntens}, \ref{eqnth} and \ref{eqnfh}, we get
\begin{align*}
[T_h(\omega&)](x_2)=\int_{\mathbb{R}} \omega(x_1)\int_{\mathbb{R}^2}f(u,v)K_h(x_1-u)K_h(x_2-v)dudv dx_1\\
&=\int_{\mathbb{R}} \omega(x_1)\int_{\mathbb{R}^2}\sum_{m=1}^{rank(T)}\sigma_m v_m(v)u_m(u)K_h(x_1-u)K_h(x_2-v)dudv dx_1\\
&=\sum_{m=1}^{rank(T)}\sigma_m \   v_m\star K_h(x_2)\int_{\mathbb{R}} \omega(x_1)u_m\star K_h(x_1)dx_1,
\end{align*}
and we conclude that 
\begin{equation}
\label{0}
T_h=\sum_{m=1}^{rank(T)}\sigma_m \   v_m\star K_h  \otimes  u_m\star K_h.
\end{equation}
Here $u_m\star K_h$ (similarly for  $v_m\star K_h$) denote the convolution $u_m$ and $K_h$ defined by 
\begin{equation*}
u_m\star K_h (x_1)=\int_{\mathbb{R}} u_m(u)K_h(x_1-u) du.
\end{equation*}
Given $\omega \in L^2(\mathbb{R})$, let ${\cal F} [\omega]$ denote its Fourier transform. We have  ${\cal F} [u_m\star K_h ]={\cal F} [u_m]{\cal F} [K_h]$ and ${\cal F} [v_m\star K_h ]={\cal F} [v_m]{\cal F} [K_h]$, and the linearity and invertibility of the Fourier transform imply that $\{u_m\star K_h \}_{m=1}^{rank(T)}$ is linearly independent if and only if $\{{\cal F} [u_m\star K_h ]\}_{m=1}^{rank(T)}$ is linearly independent. Since, by assumption, ${\cal F} [K_h]$ vanishes on a null set, the linear independence of $\{{\cal F} [u_m\star K_h ]\}_{m=1}^{rank(T)}$ is equivalent to that of $\{{\cal F} [u_m]\}_{m=1}^{rank(T)}$.  By linearity and invertibility of the Fourier transform, the functions $\{{\cal F} [u_m]\}_{m=1}^{rank(T)}$ are linearly independent since the functions $\{[u_m\}_{m=1}^{rank(T)}$ are linearly independent (they are orthonormal). We thus conclude that $\{u_m\star K_h \}_{m=1}^{rank(T)}$  and $\{v_m\star K_h \}_{m=1}^{rank(T)}$ are both sets of linearly independent functions. An argument similar to that used in the proof of Proposition \ref{prop01} then yields that the operator $T_h$ given by equation \ref{0} has rank equal to $rank(T)$.
\end{proof}
%%%%%%%%%%%%%%%%%%%%%%%%%%%%%%%%%%%%%%%%%%%%%%%%%%%%%%%%%%%%%%%%%%%%%%%%%%%%%%%%%%%%%%%%%%%%%%%%%%%%%%%%%%%%%%%%%%%%%%%%%%%%%%%%%%%%

\begin{proof}{\bf (Proof of Proposition \ref{propdiscr})}
Let $F$ denote the distribution of $X$. From equation \ref{eqn50}, we have
\begin{equation}
f_h(x_1,x_2)=\int K_h(x_1-u) K_h(x_2-v)dF(u,v).
\end{equation}
Using equation \ref{eqnmix} then yields
\begin{equation}
\label{eqndiscrp1}
f_h(x_1,x_2)=\sum_{m=1}^M \pi_m \int K_h(x_1-u) dF^1_{m}(u)\int K_h(x_2-v) dF^2_{m}(v).
\end{equation}
Note that the functions $\int K_h(x_1-u) dF^1_{m}(u)$ and $\int K_h(x_2-v) dF^2_{m}(v)$ are square integrable, hence the integral operators $T_h$ are Hilbert-Schmidt. Indeed, using Jensen and Fubini's inequality, as well as the fact that $K \in L^2(\mathbb{R})$, we have
\begin{equation}
\int_{\mathbb{R}}\left(\int K_h(x-u) dF^1_{m}(u)\right)^2 dx\leq \int_{\mathbb{R}} (K_h(x))^2 dx <\infty.
\end{equation}
Using an argument similar to the one used in the proof of Proposition \ref{prop01}, equation \ref{eqndiscrp1} implies that the operators $T_h$ have rank at most $M$. By taking the Fourier transforms of the functions $\{\int K_h(x_1-u) dF^1_{m}(u)\}_{m=1}^M$ and $\{\int K_h(x_2-v) dF^2_{m}(v)\}_{m=1}^M$, and by arguing as in the proof of Proposition \ref{prop02} (using the linearity and the invertibility of the Fourier transfrom), The dimension of the subspace spanned by $\{\int K_h(x_1-u) dF^1_{m}(u)\}_{m=1}^M$ is the same as that of the subspace spanned by $\{F^1_m\}_{m=1}^M$ (a similar statement applies to the functions  $\{\int K_h(x_2-v) dF^2_{m}(v)\}_{m=1}^M$). Hence the rank of $T_h$ is independent of $h>0$, and $rank(T_h)=M$ when $\{F^1_m\}_{m=1}^M$ and $\{F^2_m\}_{m=1}^M$ are linearly independent sets, i.e, when Assumption \ref{FR} holds.

\end{proof}
%%%%%%%%%%%%%%%%%%%%%%%%%%%%%%%%%%%%%%%%%%%%%%%%%%%%%%%%%%%%%%%%
%%%%%%%%%%%%%%%%%%%%%%%%%%%%%%%%%%%%%%%%%%%%%%%%%%%%%%%%%%%%%%%%
%%%%%%%%%%%%%%%%%%%%%%%%%%%%%%%%%%%%%%%%%%%%%%%%%%%%%%%%%%%%%%%
\begin{proof}{\bf (Proof of Proposition \ref{prop03})}
We first establish identity \ref{eqncon1} and \ref{eqncon2} under the assumption that the supports ${\cal S}_i$, $i\in \{1,2\}$, have finite Lebesgue measure, and then use an approximation argument to establish \ref{eqncon2} when the Lebesgue measure of the supports ${\cal S}_i$ is not necessarily finite.  Let $a \in \mathbb{R}^{|\Delta^1|}$, $b \in \mathbb{R}^{|\Delta^2|}$, and let $\langle \cdot,\cdot\rangle_2$ denote the inner product on $L^2({\cal S}_2)$. We have
\begin{align*}
b^T\Gamma_{\Delta^2}^* \circ T \circ \Gamma_{\Delta^1}(a) &= \langle\Gamma_{\Delta^2}(b), T \circ \Gamma_{\Delta^1}(a) \rangle_2\\
&=\sum_{i=1}^{|\Delta^1|}\sum_{j=1}^{|\Delta^2|} a_i b_j \int_{\delta_j^2}\int_{\delta_i^1} f(x_1,x_2)dx_1 dx_2\\
&=\sum_{i=1}^{|\Delta^1|}\sum_{j=1}^{|\Delta^2|} a_i b_j [P_{\Delta}]_{i,j}\\
&=a^TP_{\Delta}b,
\end{align*}
which establishes identity \ref{eqncon1}, and inequality \ref{eqncon2} is a direct consequence. We now prove that inequality \ref{eqncon2} is an equality for some partitions $\Delta$. The singular value decomposition \ref{eqntens} of the integral operator $T$ implies that $P_{\Delta}$ has the following representation (contrast to  equation \ref{matdec})
\begin{equation}
\label{eqncon3}
P_{\Delta}^T=\sum_{m=1}^{rank(T)} \sigma_m Q_2^m \otimes Q_1^m
\end{equation}
where  $Q^1_m$ (with a similar expression for $Q^2_m$) is a vector in $\mathbb{R}^{|\Delta^1|}$, with $i^{th}$ element given by $[Q^1_m]_i=\int_{\delta_i^1} u_m(x_1)dx_1$ (note that identity \ref{eqncon3} yields an alternative proof of inequality \ref{eqncon2}). Here the functions $u_m$ are the eigenfunctions of $T^*T$ that appear in the singular value decomposition \ref{matdec}. Since the functions $\{u_m\}_{m=1}^{rank(T)}$ ( resp. $\{v_m\}_{m=1}^{rank(T)}$) are orthonormal, they are necessarily linearly independent. Hence there exist partitions $\Delta^1$ (resp. $\Delta^2$) of the support of $X^1$ (resp. $X^2$) such that the vectors $\{Q^1_m\}_{m=1}^{rank(T)}$ (resp.$\{Q^2_m\}_{m=1}^{rank(T)}$) are linearly independent (see the proof of Proposition 3$-$part (a)$-$in Kasahara and Shimotsu \cite{HKKS}); it then follows by an argument similar to that used in the proof of \ref{prop01} that $rank(P_{\Delta})=rank(T)$ for such a partition $\Delta$.  We now establish inequality \ref{eqncon2} when the Lebesgue measure of the supports ${\cal S}_i$ is not necessarily finite. For $R>0$ define the matrix $P_{\Delta,R}$ by 
$$[P_{\Delta,R}]_{i,j}=P(X^1 \in \delta_i^1 \cap [-R,R]^{d_1}, X^2 \in \delta_j^2\cap [-R,R]^{d_2})$$
where $d_i$ is such that ${\cal S}_i \subset \mathbb{R}^{d_i}$, and let $T_{R}$ be defined by
$$[T_{R}(u)](x_2)=\int_{{\cal S}_1}u(x_1)f(x_1,x_2)\mathbb{I}_{\mathbb{R}^{d_1}}(x_1)\mathbb{I}_{\mathbb{R}^{d_2}}(x_2)dx_1.$$
Note that $rank (T_{R})\leq rank(T)$ (indeed, the linear independence of a subset of $\{f^i_m\mathbb{I}_{R^{d_i}}\}_{m=1}^M$ implies the linear independent of the corresponding subset of $\{f^i_m\}_{m=1}^M$), and by inequality \ref{eqnweyl}, for all sufficiently large R's, we have $rank(P_{\Delta})\leq rank(P_{\Delta,R})$. The preceding argument (for the case where the supports ${\cal S}_i$ have finite Lebesgue measure) yields $rank(P_{\Delta,R})\leq rank(T_{R})$. We thus conclude that for $R$ sufficiently large $$rank(P_{\Delta})\leq rank(P_{\Delta,R})\leq rank(T_{R})\leq rank(T).$$
\end{proof}
%%%%%%%%%%%%%%%%%%%%%%%%%%%%%%%%%%%%%%%%%%%%%%%%%%%%%%%%%%%%%%%%%%%%%%%%%%%%%%%%%%%%%%%%%%%%%%%%%%%%%%%%%%%%%%%%%%%%%%%%%%%%%%%%%%%%
\begin{proof}{\bf (Proof of Proposition \ref{prop50})}
We first establish inequality \ref{eqnconc1}. It follows from a slight modification of Theorem $3.4$ of Pinelis \cite{PIN}, applied to sums of independent random elements in the space of Hilbert-Schmidt operators on $L^2(\mathbb{R})$ (which is a $(2,1)$-smooth space in the setting of Pinelis \cite{PIN}).
Let $\eta_i$ be defined by $\eta_i=T_{h,X_i}-ET_{h,X}$, and note that  if $X'$ is an independent copy of $X$, then we have $\|\eta_i\|_{HS}\leq E \|T_{h,X}-T_{h,X'}\|_{HS} \leq L_h$ (with $T_{h,X_i}$ defined as in equation \ref{eqn050}).
By Theorem 3.2 of Pinelis \cite{PIN}, for all $\lambda>0$ and $\tau>0$, we have
\begin{equation}
\label{eqnpin1}
P(\|\hat{T}_h-T_h\|_{HS}>\tau)\leq 2 \exp(N \tau \lambda) (1+u)^N,
\end{equation}
where $u=E\left[e^{\lambda \|\eta_1\|_{HS}}-1-\lambda \|\eta_1\|_{HS}\right]$. Following the same steps as the begining of the proof of Theorem $3.4$ in Pinelis \cite{PIN}, we get:
\begin{equation}
\label{eqnpin2}
P(\|\hat{T}_h-T_h\|_{HS}>\tau)\leq 2 \exp\left(-\lambda \tau N+N \ln\left(1+\frac{e^{\lambda L_h}-1-\lambda L_h}{L_h^2}\sigma_h^2\right)\right)
\end{equation}
for all $\tau,\lambda>0$, and where $\sigma_h^2$ is as defined in Proposition \ref{prop50}. It then remains to make an appropriate choice for $\lambda>0$.  Note that the optimal value of $\lambda$ that minimizes the right hand side of inequality \ref{eqnpin2} cannot be obtained in closed form. As in Pinelis \cite{PIN}, we choose for $\lambda$ the value that minimizes the right hand side of the inequality 
\begin{equation}
\label{eqnpin3}
P(\|\hat{T}_h-T_h\|_{HS}>\tau)\leq 2 \exp\left(-\lambda \tau N+N \frac{e^{\lambda L_h}-1-\lambda L_h}{L_h^2}\sigma_h^2\right)
\end{equation}
obtained from inequality \ref{eqnpin2} by using the inequality $\ln(1+x)\leq x$, for all $x>0$. That value of $\lambda$ is given by 
\begin{equation}
\label{eqnpin4}
\lambda*=\frac{1}{L_h}\ln\left(1+\frac{\tau L_h }{\sigma_h^2}\right).
\end{equation}
Inequality \ref{eqnconc1} is obtained by substituting this value of $\lambda$ into inequality \ref{eqnpin2}. Note that inequality 3.4 in Pinelis \cite{PIN} would correspond to substituting $\lambda*$ into inequality \ref{eqnpin3}. However, since the right hand side of inequality \ref{eqnpin3} is larger than the right-hand side of\ref{eqnpin2}, the right hand side of inequality 3.4 in Pinelis \cite{PIN} is larger than the right hand side of inequality \ref{eqnconc1}, and this difference leads to noticeable improvement in the performance of our procedure (especially for small values of $N$). \\
To establish inequality \ref{eqn52}, we use lemma 2 in Smale and Zhou \cite{SZ}, also derived from Theorem 3.4 of Pinelis \cite{PIN}, which (for all $0<\delta<1$) yields 
\begin{equation}
\label{eqnpin5}
\|\hat{T}_h-T_h\|_{HS} =\|\frac{1}{N}\sum_{i=1}^N \eta_i \|_{HS} \leq \frac{2 L_h ln(2/\delta))}{N}+\sqrt{\frac{2ln(2/\delta) E\|\eta_i\|_{HS}^2}{N}}
\end{equation} 
with probability greater than $1-\delta$. Note that given $X'$, an independent copy of $X$, we have
 \begin{align*}
  & E\| T_{h,X}-T_{h,X'} \|_{HS}^2 \\
%  &= E \int_{\mathbb{R}^2} \left(K_h(X_1-x_1)K_h(X_2-x_2)-K_h(X_1'-x_1)K_h(X_2'-x_2)\right)^2dx_1dx_2\\
%  &= \int_{\mathbb{R}^2} E\left(K_h(X_1-x_1)K_h(X_2-x_2)-K_h(X_1'-x_1)K_h(X_2'-x_2)\right)^2dx_1dx_2\\
% &=2 \int_{\mathbb{R}^2} E\left(K_h(X_1-x_1)K_h(X_2-x_2)-E\{K_h(X_1-x_1)K_h(X_2-x_2)\}\right)^2dx_1dx_2\\
  &=2 E \int_{\mathbb{R}^2} \left(K_h(X_1-x_1)K_h(X_2-x_2)-E\{K_h(X_1-x_1)K_h(X_2-x_2)\}\right)^2dx\\
    &=2 E \|T_{h,X}-ET_{h,X}\|_{HS}^2=2 E\|\eta_i\|_{HS}^2.\\
 \end{align*}
 
\indent Inequality \ref{eqn52} then follows from combining inequality \ref{eqnpin5} with Hoeffding's concentration inequality (for U-statistics Hoeffding \cite{WH}), which yields (for all $0<\delta<1$):
\begin{equation}
\label{eqnhoeff}
E \|T_{h,X}-T_{h,X'}\|_{HS}^2 \leq \frac{1}{N(N-1)}\sum_{i\neq j} \|T_{h,X_i}-T_{h,X_j}\|_{HS}^2+ L_h^2 \sqrt{\frac{ln(1/\delta)}{N}}
\end{equation}
with probability greater than $1-\delta$.

\end{proof}
%%%%%%%%%%%%%%%%%%%%%%%%%%%%%%%%%%%%%%%%%%%%%%%%%%%%%%%%%%%%%%%%%%%%%%%%%%%%%%%%%%%%%%%%%%%%%%%%%%%%%%%%%%%%%%%%%%%%%%%%%%%%%%%%%%%%
\begin{proof}{\bf (Proof of Proposition \ref{prop1})}

%By the definition of the operator norm, we have:
%\begin{equation*}
%\|\hat{T}-T\|=\sup_{\left\{\|w\|_{L^2(\mathcal{ S}_1)} \leq 1\right\}} \left[ \int_{\mathcal{ S}_2}\left(\int_{\mathcal{ S}_1} w(x_1)(f(x_1,x_2)-\hat{f}(x_1,x_2))dx_1\right)^2dx_2\right]^{1/2}.
%\end{equation*}
%Using Minkowski's integral inequality, the left-hand side is bounded by:
% \begin{equation*}
%\|\hat{T}-T\|\leq \sup_{\left\{\|w\|_{L^2(\mathcal{ S}_1)} \leq 1\right\}} \int_{\mathcal{ S}_1}|w(x_1)|\left(\int_{\mathcal{ S}_2}(f(x_1,x_2)-\hat{f}(x_1,x_2))^2dx_2\right)^{1/2}dx_1.
%\end{equation*}

We first establish inequality \ref{eqnh0}; it follows from properties of convolutions. Let $G:\mathbb{R}^2\rightarrow \mathbb{R}$ be defined by $G(x_1,x_2)=K(x_1)K(x_2)$, where $K$ is the kernel in Proposition \ref{prop1}. By Fubini's theorem, the assumption on $K$ imply that: $G\in L^1(\mathbb{R}^2)\cap  L^2(\mathbb{R}^2)$, and $\int_{\mathbb{R}^2}G=1$. For $h>0$, let $G_h$ be defined by $G_h(x)=1/h^2G(x/h)$, for all $x \in \mathbb{R}^2$. Note that $G_h(x)=K_h(x_1)K_h(x_2)$ and that $f_h=f\star G_h$. From $\int_{\mathbb{R}^2}G_h=1$, we get
\begin{align*}
\|T_h-T\|_{HS}^2&=\int_{\mathbb{R}^2}(f_h(x)-f(x))^2dx\\
&=\int_{\mathbb{R}^2}\left(\int_{\mathbb{R}^2}(f(x-hy)-f(y))G(y)dy\right)^2dx.
\end{align*}
Letting $\alpha:=\int_{\mathbb{R}^2}|G|$, and using Jensen and (then) Fubini's inequality yields
\begin{align*}
\|T_h-T\|_{HS}^2&\leq \alpha \int_{\mathbb{R}^2}|G(y)|\int_{\mathbb{R}^2}|f(x-hy)-f(x)|^2dx dy.
\end{align*}
The integrand $|G(y)|\int_{\mathbb{R}^2}|f(x-hy)-f(x)|^2dx$ is dominated by $4\|f\|_2^2|G(y)|$, and by Lemma 4.3 of Brezis \cite{HB} converges pointwise to zero as $h\rightarrow 0$. Hence, by the Dominated Convergence Theorem, we get 
\begin{equation*}
\lim_{h\rightarrow 0}\|T-T_h\|_{HS}=0.
\end{equation*}
The above argument is a modification of the proof of Theorem 4.22 in Brezis \cite{HB}, to account for the fact that the function $G$ is not necessarily a ``mollifier". To establish the second part of inequality \ref{eqnh0}, we proceed similarly. 
For  $\epsilon>0$ and small, let $f_{\epsilon}\in C_c(\mathbb{R}^2)$ be such that $\|f-f_{\epsilon}\|_2\leq \epsilon$ (recall that $C_c(\mathbb{R}^2)$ is dense in $L^2(\mathbb{R}^2)$). We have
\begin{equation*}
\|T_h\|_{HS} \leq \|G_h\star (f-f_{\epsilon})\|_2+\|G_h\star f_{\epsilon}\|_2.
\end{equation*}
Using Young's convolution inequality (Theorem 4.15 in Brezis \cite{HB}), we get
\begin{equation*}
\|T_h\|_{HS} \leq \epsilon \|G\|_1+\|G_h\|_2  \|f_{\epsilon}\|_1.
\end{equation*}
The second term on the right-hand side converges to zero as $h \rightarrow \infty$, since $\|G_h\|_2=O(1/h)$. As $\epsilon>0$ is arbitrary, we conclude that 
\begin{equation*}
\lim_{h\rightarrow \infty} \|T_h\|_{HS}=0.
\end{equation*}

 We now establish inequality \ref{eqnsimt}. By the definition of the operator norm and by Cauchy-Schwartz inequality, we have
\begin{equation*}
\|\hat{T}-T\|_{op}^2\leq \|\hat{T}-T\|_{HS}^2 =\int_{\mathbb{R}^2}(f(x_1,x_2)-\hat{f}_h(x_1,x_2))^2dx.
\end{equation*}
Hence
\begin{equation}
\label{Ceqn001}
E\|\hat{T}-T\|_{op}^2\leq \int_{\mathbb{R}^2}E(f(x_1,x_2)-\hat{f}_h(x_1,x_2))^2 dx.
\end{equation}
The right-hand side of inequality \ref{Ceqn001} represents the integrated mean-squared error (IMSE) of the estimator of the density $\hat{f}$, which by a standard argument decomposes into a bias and variance term:
\begin{equation*}
\int_{\mathbb{R}^2}E(f(x_1,x_2)-\hat{f}_h(x_1,x_2))^2 dx=\|T_h-T\|_{HS}^2+E \int_{\mathbb{R}^2}(\hat{f}_h(x)-f_h(x))^2 dx.
\end{equation*}
By the first part of inequality \ref{eqnh0}, the bias term converges to zero as $h\rightarrow 0$, and a standard computation shows that the variance term converges to zero if $N h^2\rightarrow \infty$. Inequality \ref{eqnsimt} then follows from Jensen's inequality.
\end{proof}
%%%%%%%%%%%%%%%%%%%%%%%%%%%%%%%%%%%%%%%%%%%%%%%%%%%%%%%%%%%%%%%%%%%%%%%%%%%%%%%%%%%%%%%%%%%%%%%%%%%%%%%%%%%%%%%%%%%%%%%%%%%%%%%%%%%%%
\begin{proof}{\bf (Proof of Theorem \ref{theo1})}
Note that under the mixture representation \ref{eqnmix}, the singular values of $T$ satisfy: $\sigma_R(T)>0$ and $\sigma_{R+1}(T)=0$, where $R$ denotes the rank of $T$. Also, by inequality \ref{eqnHW} and the triangle inequality, for all $j\in \{1,\cdots,N\}$, we have
\begin{equation}
\label{eqn000}
|r_j(\hat{T}_h)-r_j(T_h)| \leq \|T_h-\hat{T}_h\|_{HS},
\end{equation}
 where $r_j(\hat{T}_h)$ is defined as in equation \ref{eqnr}. Given the result in Proposition \ref{prop50}, to establish inequality \ref{eqnest1}, it suffices to show that $\{\|\hat{T}_h-T_h\|_{HS}\leq \hat{\tau}_h(N,\delta)\}\subset\{  r_{R+1}(\hat{T}_h)<\hat{\tau}_h(N,\delta)\}$.  The latter is a direct consequence of  inequality \ref{eqn000}, as $r_{R+1}(\hat{T}_h) \leq r_{R+1}(T_h)+\|T_h-\hat{T}_h\|_{HS}$ and $ r_{R+1}(T_h)=0$. To establish inequality \ref{eqnest2}, it suffices to show that
\begin{equation}
 \label{eqn001}
 \begin{aligned}
\{ \sigma_R(T_h) > 2\hat{\tau}_h(N,\delta)\} \cap \{\|\hat{T}_h-T_h\|_{HS}\leq \hat{\tau}_h(N,\delta)\} \\ \subset \{r_{R}(\hat{T}_h) \geq \hat{\tau}_h(N,\delta\} \cap \{r_{R+1}(\hat{T}_h)<\hat{\tau}_h(N,\delta)\} .\end{aligned}
\end{equation}
 From inequality \ref{eqn000}, we have $r_R(\hat{T}_h)\geq r_R(T_h)-\|T_h-\hat{T}_h\|_{HS}= \sigma_R(T_h) -\|T_h-\hat{T}_h\|_{HS}\geq  \hat{\tau}_h(N,\delta)$ on the event $\{ \sigma_R(T_h) > 2\hat{\tau}_h(N,\delta)\} \cap \{\|\hat{T}_h-T_h\_{HS}|\leq \hat{\tau}_h(N,\delta)\} $. In addition, as in the proof of inequality \ref{eqnest1}, $r_{R+1}(\hat{T}_h)<\hat{\tau}_h(N,\delta)$ on the event $ \{\|\hat{T}_h-T_h\|_{HS}\leq \hat{\tau}_h(N,\delta)\} $. Therefore, the inclusion \ref{eqn001} holds, and inequality \ref{eqnest2} follows. Finally, to establish inequality \ref{eqnest3}, it suffices to verify the inclusion
 \begin{equation}
 \label{Ceqn002}
 \{ \sigma_{R}(T_h)+\|\hat{T}_h-T_h\|_{HS}<\hat{\tau}_h(N,\delta)\}\subset  \{r_{R}(\hat{T}_h) < \hat{\tau}_h(N,\delta\}\},
 \end{equation}
which follows from inequality \ref{eqn000}, as $r_{R}(\hat{T}_h)\leq r_{R}(T_h)+\|T_h-\hat{T}_h\|_{HS}= \sigma_{R}(T_h)+\|T_h-\hat{T}_h\|_{HS}$.

\end{proof}
%%%%%%%%%%%%%%%%%%%%%%%%%%%%%%%%%%%%%%%%%%%%%%%%%%%%%%%%%%%%%%%%%%%%%%
\begin{proof}{\bf (Proof of Corollary \ref{cor1})}
Given the random sample $\{X_{i}\}_{i=1}^N$, define the random vector spaces ${\cal \hat{H}}_j$, $j\in \{1,2\}$, by
\begin{equation*} 
{\cal \hat{H}}_j=span\{K_h(X^j_{i}-\cdot)| i=1,\cdots, N\}.
\end{equation*}
Note that the operator $\hat{T}_h$ has range in ${\cal \hat{H}}_2$. Indeed, for $w \in L^2({\cal S}_1)$, we have
\begin{equation}
\label{eqnp0}
[\hat{T}_h(w)](x_2)=(1/N)\sum_{i=1}^N K_h(X^2_{i}-y) \int_{{\cal S}_1} w(x)K_h(X^1_{i}-x_1)dx_1.
\end{equation}
Moreover, since the kernel $K$ has a non-vanishing Fourier transform, the vector spaces ${\cal \hat{H}}_1$ and ${\cal \hat{H}}_2$ have dimension equal to $N$, as long as the $X_{1i}$'s and the $X_{2i}$'s are all distinct, and the latter occurs with probability one  (it can be easily shown$-$by considering their Fourier transforms$-$that the functions $\{K_h(X^1_{i}-\cdot)\}_{i=1}^N$ are linearly independent if the $X_{1i}$'s are distinct). Let $\Gamma_1:\mathbb{R}^N\rightarrow {\cal \hat{H}}_1$ and $\Gamma_2:\mathbb{R}^N\rightarrow {\cal \hat{H}}_2$ be defined by: 
\begin{equation*}
\Gamma_1(a)=\sum_{i=1}^N a_i K_h(X^1_{i}-\cdot)
\end{equation*}
and 
\begin{equation*}
\Gamma_2(a)=\sum_{i=1}^N a_i K_h(X^2_{i}-\cdot),
\end{equation*}
where $a \in \mathbb{R}^N$.
Note that 
\begin{equation}
\label{eqnp1}
\|\Gamma_1(a)\|^2_{L^2({\cal S}_1)}=a^T \hat{W}_{1h} a \ \ \text{and}\ \ \|\Gamma_2(a)\|^2_{L^2({\cal S}_2)}=a^T \hat{W}_{2h} a
\end{equation}
where the matrices $\hat{W}_{1h}$ and $\hat{W}_{2h}$ are as defined in equation \ref{eqn02}. Since the matrices $\hat{W}_{1h}$ and $\hat{W}_{2h}$ are symmetric and positive definite (see equation \ref{eqn02}, and recall that the functions  $\{K_h(X^1_{i}-\cdot)\}_{i=1}^N$ are linearly independent with probability one), their powers $(\hat{W}_{1h})^{d}$ and $(\hat{W}_{2h})^{d}$, for any $d \in \mathbb{R}$, are well defined. 
Let $R:\mathbb{R}^N\rightarrow {\cal \hat{H}}_1$ and $S:{\cal \hat{H}}_2 \rightarrow \mathbb{R}^N$ be defined by 
\begin{equation*}
Ra=\Gamma_1(\hat{W}_{1h}^{-1/2}a) \ \ \text{and} \ \ S(\Gamma_2(a))=\hat{W}_{2h}^{1/2} a.
\end{equation*}
It follows from equation \ref{eqnp1} that the operators $S$ and $R$ are isometries, i.e, $\|Ra\|_{L^2({\cal S}_1)}=\|a\|$ and $\|S(\Gamma_2(a))\|=\|\Gamma_2(a)\|_{L^2({\cal S}_2)}$. Also, using the representation of equation \ref{eqnp0}, it can be shown that 
\begin{equation}
\label{eqnp2}
\hat{T}_h(\Gamma_1(a))=(1/N)\Gamma_2(W_1 a)
\end{equation}
 Let $\langle\cdot,\cdot\rangle$ denote the inner product on $\mathbb{R}^N$. We show below that the operator $\tilde{T}_h=S\hat{T}_hR:\mathbb{R}^N\rightarrow \mathbb{R}^N$ has the same singular values as $\hat{T}_h$. Moreover, the matrix representation of the operator $\tilde{T}_h$ is given by $\hat{A}_h$ in equation \ref{eqn01}. Indeed, for $a,b\in \mathbb{R}^N$, identity \ref{eqnp2} yields
\begin{align*}
\langle b, \tilde{T}_ha\rangle &= \langle b, S \hat{T}_h R a\rangle \\
&= \langle b, S \hat{T}_h \Gamma_1 (\hat{W}_{1h}^{-1/2}a) \rangle\\
&= \langle b, (1/N) S \Gamma_2(\hat{W}_{1h}^{1/2}a) \rangle\\
&=\langle b, (1/N) \hat{W}_{2h}^{1/2}\hat{W}_{1h}^{1/2} a\rangle=\langle b, \hat{A}_h a\rangle.
\end{align*}
It now remains to show that $\tilde{T}_h$ and $\hat{T}_h$ have the same singular values. This follows by noting that given a singular value decomposition 
\begin{equation*}
\hat{T}_h=\sum_{i=1}^N \sigma_i(\hat{T}_h) \hat{v}_i \otimes \hat{u}_i
\end{equation*}
of $\hat{T}_h$, the operator $\tilde{T}_h$ has the representation
\begin{equation}
\label{eqn002}
\tilde{T}_h=\sum_{i=1}^N \sigma_i(\hat{T}_h) S\hat{v}_i \otimes R^*\hat{u}_i,
\end{equation}
where $R^*$ denotes the adjoint of $R$. Since the sets $\{S\hat{v}_i \}_{i=1}^N$ and $\{R^*\hat{u}_i\}_{i=1}^N$ are orthonormal (R and S are isometries), \ref{eqn002} represents a singular value decomposition of $\tilde{T}_h$, and we conclude that $\tilde{T}_h$ and $\hat{T}_h$ have the same singular values.
\end{proof}

\section*{Acknowledgements}
We thank Ivan Canay, Denis Chetverikov and Joel Horowitz for their helpful comments and suggestions. we would  also like to thank the Editor, an Associate Editor, and two anonymous Referees for a careful reading of the manuscript and for comments that greatly improved this paper. We are also grateful to Hiro Kasahara and Katsumi Shimotsu for providing us with their code and datasets for the empirical section of this paper.
% \begin{supplement}
% \sname{Supplement A}\label{suppA}
% \stitle{Title of the Supplement A}
% \slink[url]{http://www.e-publications.org/ims/support/dowload/imsart-ims.zip}
% \sdescription{Dum esset rex in
% accubitu suo, nardus mea dedit odorem suavitatis. Quoniam confortavit
% seras portarum tuarum, benedixit filiis tuis in te. Qui posuit fines tuos}
% \end{supplement}

% 
% \begin{thebibliography}{9}
% 
% \bibitem{CMN}
% Compiani, G. and Kitamura, Y.(2016). \textit{Using Mixtures in Econometric Models: A Brief Review and Some New Results}, 2nd ed.
% Wiley, New York.
% \MR{1700749}
% 
% 
% 
% 
% \bibitem{r2}
% \textsc{Bourbaki, N.}  (1966). \textit{General Topology}  \textbf{1}.
% Addison--Wesley, Reading, MA.
% 
% \bibitem{r3}
% \textsc{Ethier, S. N.} and \textsc{Kurtz, T. G.} (1985).
% \textit{Markov Processes: Characterization and Convergence}.
% Wiley, New York.
% \MR{838085}
% 
% \bibitem{r4}
% \textsc{Prokhorov, Yu.} (1956).
% Convergence of random processes and limit theorems in probability
% theory. \textit{Theory  Probab.  Appl.}
% \textbf{1} 157--214.
% \MR{84896}
% 
% \bibitem{CK}
% \textsc{Compiani, G. and Kitamura, Y.} (2016).
% Using Mixtures in Econometric Models: A Brief Review and Some New Results.}},
% \textit{Econometrics Journal}
% \textbf{19-3} 95 -- 127.
% 
% 
% \end{thebibliography}

\bibliographystyle{imsart-nameyear}
\bibliography{biblio}

\end{document}

%% file: Tables/design1.tex
\begin{tabular}{cc|cccc|cccc}
\hline 
 & \multicolumn{1}{c}{} & \multicolumn{4}{c}{$N=500$} & \multicolumn{4}{c}{$N=2000$}\tabularnewline
\hline 
\multicolumn{2}{c|}{Method} & $M=1$ & $M=2$ & $M=3$ & $M\protect\geq4$ & $M=1$ & $M=2$ & $M=3$ & $M\protect\geq4$\tabularnewline
\hline 
\multicolumn{2}{c|}{SVT} & 0.035 & 0.961 & 0.004 & 0.000 & 0.000 & 0.606 & 0.394 & 0.000\tabularnewline
\hline 
$M_{0}=4$ & SHT & 0.021 & 0.891 & 0.082 & 0.006 & 0.000 & 0.566 & 0.414 & 0.020\tabularnewline
 & AIC & 0.004 & 0.757 & 0.215 & 0.024 & 0.000 & 0.317 & 0.609 & 0.074\tabularnewline
 & BIC & 0.464 & 0.533 & 0.003 & 0.000 & 0.000 & 0.989 & 0.011 & 0.000\tabularnewline
 & HQ & 0.092 & 0.876 & 0.031 & 0.001 & 0.000 & 0.766 & 0.226 & 0.008\tabularnewline
\hline 
$M_{0}=8$ & SHT & 0.094 & 0.874 & 0.032 & 0.000 & 0.000 & 0.690 & 0.306 & 0.004\tabularnewline
 & AIC & 0.022 & 0.830 & 0.148 & 0.000 & 0.000 & 0.384 & 0.542 & 0.074\tabularnewline
 & BIC & 0.704 & 0.296 & 0.000 & 0.000 & 0.000 & 1.000 & 0.000 & 0.000\tabularnewline
 & HQ & 0.212 & 0.788 & 0.000 & 0.000 & 0.000 & 0.954 & 0.046 & 0.000\tabularnewline
\end{tabular}

%% file: Tables/design2.tex
\begin{tabular}{cc|cccc|cccc}
\hline 
 & \multicolumn{1}{c}{} & \multicolumn{4}{c}{$N=$500} & \multicolumn{4}{c}{$N=2000$}\tabularnewline
\hline 
\multicolumn{2}{c|}{Method} & $M=1$ & $M=2$ & $M=3$ & $M\protect\geq4$ & $M=1$ & $M=2$ & $M=3$ & $M\protect\geq4$\tabularnewline
\hline 
\multicolumn{2}{c|}{SVT} & 0.000 & 0.000 & 1.000 & 0.000 & 0.000 & 0.000 & 1.000 & 0.000\tabularnewline
\hline 
$M_{0}=4$ & SHT & 0.425 & 0.000 & 0.575 & 0.000 & 0.520 & 0.000 & 0.480 & 0.000\tabularnewline
 & AIC & 0.454 & 0.000 & 0.544 & 0.002 & 0.452 & 0.000 & 0.492 & 0.056\tabularnewline
 & BIC & 0.410 & 0.000 & 0.590 & 0.000 & 0.497 & 0.000 & 0.458 & 0.045\tabularnewline
 & HQ & 0.422 & 0.000 & 0.578 & 0.000 & 0.520 & 0.000 & 0.462 & 0.018\tabularnewline
\hline 
$M_{0}=8$ & SHT & 0.382 & 0.013 & 0.082 & 0.523 & 0.478 & 0.244 & 0.002 & 0.276\tabularnewline
 & AIC & 0.362 & 0.018 & 0.028 & 0.592 & 0.466 & 0.204 & 0.000 & 0.330\tabularnewline
 & BIC & 0.339 & 0.028 & 0.140 & 0.493 & 0.472 & 0.224 & 0.004 & 0.300\tabularnewline
 & HQ & 0.352 & 0.018 & 0.076 & 0.554 & 0.476 & 0.282 & 0.000 & 0.242\tabularnewline
\end{tabular}

%% file: Tables/design3.tex
\begin{tabular}{cc|cccc|cccc}
\hline 
 & \multicolumn{1}{c}{} & \multicolumn{4}{c}{$N=500$} & \multicolumn{4}{c}{$N=2000$}\tabularnewline
\hline 
\multicolumn{2}{c|}{Method} & $M=1$ & $M=2$ & $M=3$ & $M\protect\geq4$ & $M=1$ & $M=2$ & $M=3$ & $M\protect\geq4$\tabularnewline
\hline 
\multicolumn{2}{c|}{SVT} & 0.000 & 0.000 & 1.000 & 0.000 & 0.000 & 0.000 & 1.000 & 0.000\tabularnewline
\hline 
$M_{0}=4$ & SHT & 0.000 & 0.000 & 0.980 & 0.020 & 0.000 & 0.000 & 0.950 & 0.050\tabularnewline
 & AIC & 0.000 & 0.000 & 0.886 & 0.114 & 0.000 & 0.000 & 0.882 & 0.118\tabularnewline
 & BIC & 0.000 & 0.000 & 1.000 & 0.000 & 0.000 & 0.000 & 0.992 & 0.008\tabularnewline
 & HQ & 0.000 & 0.000 & 0.978 & 0.022 & 0.000 & 0.000 & 0.958 & 0.042\tabularnewline
\hline 
$M_{0}=8$ & SHT & 0.000 & 0.000 & 0.940 & 0.060 & 0.000 & 0.000 & 0.930 & 0.070\tabularnewline
 & AIC & 0.000 & 0.000 & 0.824 & 0.176 & 0.000 & 0.000 & 0.806 & 0.194\tabularnewline
 & BIC & 0.000 & 0.000 & 0.992 & 0.008 & 0.000 & 0.000 & 0.998 & 0.002\tabularnewline
 & HQ & 0.000 & 0.000 & 0.964 & 0.036 & 0.000 & 0.000 & 0.968 & 0.032\tabularnewline
\end{tabular}

%% file: Tables/design4.tex
\begin{tabular}{cc|cccc|cccc}
\hline 
 & \multicolumn{1}{c}{} & \multicolumn{4}{c}{$N=500$} & \multicolumn{4}{c}{$N=2000$}\tabularnewline
\hline 
\multicolumn{2}{c|}{Method} & $M\protect\leq3$ & $M=4$ & $M=5$ & $M\protect\geq6$ & $M\protect\leq3$ & $M=4$ & $M=5$ & $M\protect\geq6$\tabularnewline
\hline 
\multicolumn{2}{c|}{SVT} & 0.000 & 0.066 & 0.934 & 0.000 & 0.000 & 0.000 & 1.000 & 0.000\tabularnewline
\hline 
$M_{0}=4$ & SHT & 0.466 & 0.534 & 0.000 & 0.000 & 0.484 & 0.516 & 0.000 & 0.000\tabularnewline
 & AIC & 0.475 & 0.525 & 0.000 & 0.000 & 0.478 & 0.522 & 0.000 & 0.000\tabularnewline
 & BIC & 0.481 & 0.519 & 0.000 & 0.000 & 0.470 & 0.530 & 0.000 & 0.000\tabularnewline
 & HQ & 0.480 & 0.520 & 0.000 & 0.000 & 0.482 & 0.518 & 0.000 & 0.000\tabularnewline
\hline 
$M_{0}=8$ & SHT & 0.679 & 0.170 & 0.083 & 0.068 & 0.778 & 0.072 & 0.097 & 0.053\tabularnewline
 & AIC & 0.656 & 0.213 & 0.075 & 0.056 & 0.786 & 0.078 & 0.089 & 0.047\tabularnewline
 & BIC & 0.659 & 0.227 & 0.067 & 0.047 & 0.774 & 0.109 & 0.086 & 0.031\tabularnewline
 & HQ & 0.660 & 0.216 & 0.081 & 0.043 & 0.785 & 0.093 & 0.094 & 0.028\tabularnewline
\end{tabular}

%% file: Tables/design6.tex
\begin{tabular}{cc|cccc|cccc}
\hline 
 & \multicolumn{1}{c}{} & \multicolumn{4}{c}{$N=500$} & \multicolumn{4}{c}{$N=2000$}\tabularnewline
\hline 
\multicolumn{2}{c|}{Method} & $M=1$ & $M=2$ & $M=3$ & $M\protect\geq4$ & $M=1$ & $M=2$ & $M=3$ & $M\protect\geq4$\tabularnewline
\hline 
\multicolumn{2}{c|}{SVT} & 0.000 & 0.992 & 0.008 & 0.000 & 0.000 & 0.493 & 0.507 & 0.000\tabularnewline
\hline 
\multicolumn{2}{c|}{ave-rk $(\alpha=0.05)$} & 0.142 & 0.810 & 0.047 & 0.001 & 0.005 & 0.776 & 0.214 & 0.005\tabularnewline
\multicolumn{2}{c|}{AIC by ave-rk} & 0.012 & 0.867 & 0.119 & 0.003 & 0.000 & 0.587 & 0.399 & 0.013\tabularnewline
\multicolumn{2}{c|}{BIC by ave-rk} & 0.284 & 0.715 & 0.001 & 0.000 & 0.035 & 0.942 & 0.023 & 0.000\tabularnewline
\multicolumn{2}{c|}{HQ by ave-rk} & 0.078 & 0.909 & 0.013 & 0.000 & 0.004 & 0.878 & 0.117 & 0.001\tabularnewline
\end{tabular}

%% file: Tables/design8.tex
\begin{tabular}{c||c|cccc|cccc}
\multicolumn{2}{c}{} & \multicolumn{4}{c}{$N=500$} & \multicolumn{4}{c}{$N=2000$}\tabularnewline
\hline
\multicolumn{2}{c|}{$\delta$} & $M=1$ & $M=2$ & $M=3$ & $M\protect\geq4$ & $M=1$ & $M=2$ & $M=3$ & $M\protect\geq4$\tabularnewline
\hline
\multicolumn{2}{c|}{0.05} & 0.035 & 0.961 & 0.004 & 0.000 & 0.000 & 0.606 & 0.394 & 0.000\tabularnewline
\multicolumn{2}{c|}{0.10} & 0.017 & 0.966 & 0.017 & 0.000 & 0.000 & 0.510 & 0.490 & 0.000\tabularnewline
\multicolumn{2}{c|}{0.15} & 0.003 & 0.973 & 0.024 & 0.000 & 0.000 & 0.386 & 0.614 & 0.000\tabularnewline
\multicolumn{2}{c|}{0.20} & 0.003 & 0.957 & 0.040 & 0.000 & 0.000 & 0.347 & 0.653 & 0.000\tabularnewline
\multicolumn{2}{c|}{0.25} & 0.004 & 0.950 & 0.046 & 0.000 & 0.000 & 0.279 & 0.720 & 0.001\tabularnewline
\multicolumn{2}{c|}{0.30} & 0.003 & 0.927 & 0.070 & 0.000 & 0.000 & 0.253 & 0.747 & 0.000\tabularnewline
\multicolumn{2}{c|}{0.35} & 0.001 & 0.906 & 0.093 & 0.000 & 0.000 & 0.192 & 0.807 & 0.001\tabularnewline
\multicolumn{2}{c|}{0.40} & 0.001 & 0.893 & 0.106 & 0.000 & 0.000 & 0.179 & 0.818 & 0.003\tabularnewline
\multicolumn{2}{c|}{0.45} & 0.000 & 0.886 & 0.114 & 0.000 & 0.000 & 0.149 & 0.847 & 0.004\tabularnewline
\multicolumn{2}{c|}{0.50} & 0.000 & 0.858 & 0.142 & 0.000 & 0.000 & 0.130 & 0.862 & 0.008\tabularnewline
\end{tabular}

%% file: Tables/delta.tex
\begin{tabular}{c||c|cccc|cccc}
\hline 
\multicolumn{2}{c}{} & \multicolumn{4}{c}{$N=500$} & \multicolumn{4}{c}{$N=2000$}\tabularnewline
\hline 
\multicolumn{2}{c|}{} & $M=1$ & $M=2$ & $M=3$ & $M\protect\geq4$ & $M=1$ & $M=2$ & $M=3$ & $M\protect\geq4$\tabularnewline
\hline 
\multicolumn{2}{c|}{Design 1} & 0.001 & 0.893 & 0.106 & 0.000 & 0.000 & 0.179 & 0.818 & 0.003\tabularnewline
\multicolumn{2}{c|}{Design 2} & 0.000 & 0.000 & 1.000 & 0.000 & 0.000 & 0.000 & 0.972 & 0.028\tabularnewline
\multicolumn{2}{c|}{Design 3} & 0.000 & 0.000 & 1.000 & 0.000 & 0.000 & 0.000 & 0.999 & 0.001\tabularnewline
\multicolumn{2}{c|}{Design 5} & 0.000 & 0.781 & 0.219 & 0.000 & 0.000 & 0.000 & 0.989 & 0.011\tabularnewline
\hline 
\multicolumn{2}{c|}{} & $M\protect\leq3$ & $M=4$ & $M=5$ & $M=6$ & $M\protect\leq3$ & $M=4$ & $M=5$ & $M=6$\tabularnewline
\hline 
\multicolumn{2}{c|}{Design 4} & 0.000 & 0.000 & 1.000 & 0.000 & 0.000 & 0.000 & 1.000 & 0.000\tabularnewline
\end{tabular}